\documentclass[%
aps,
prx,
superscriptaddress,
twocolumn,
nofootinbib,
noeprint,
longbibliography,
showkeys,
amsmath,amssymb,
floatfix
]{revtex4-2}

\usepackage[UKenglish]{isodate}
\usepackage[english]{babel}
\usepackage[T1]{fontenc}
\usepackage{lmodern}
\usepackage{microtype}
\usepackage{hyphenat}
\usepackage{physics}
\usepackage{bm}
\usepackage{bbold}
\usepackage{slashed}

\newcommand*\diff{\mathop{}\!\mathrm{d}}
\newcommand*\Diff[1]{\mathop{}\!\mathrm{d^{#1}}}

\newcommand*\parderiv[2]{\frac{\partial #1}{\partial #2}}
\newcommand*\lownat[1]{[#1]}
\newcommand{\bigO}{\mathcal{O}}
\newcommand{\sgn}[1]{\mathrm{sgn}(#1)}
\newcommand{\sgnb}[1]{\mathrm{sgn}_\mathrm{b}(#1)}

\usepackage{comment}
\usepackage[normalem]{ulem} 
\usepackage[dvipsnames]{xcolor}
\definecolor{darkorange}{RGB}{227, 103, 0}
\definecolor{lightblue}{RGB}{93,197, 192}
\definecolor{darkblue}{RGB}{58,129, 158}
\newcommand{\orange}[1]{\textcolor{darkorange}{#1}}
\newcommand{\blue}[1]{\textcolor{lightblue}{#1}}
\newcommand{\darkblue}[1]{\textcolor{darkblue}{#1}}

\usepackage{graphicx}
\graphicspath{{img/}}
\usepackage[caption=false]{subfig}
\usepackage{adjustbox}
\usepackage{tikz}
\usepackage{ragged2e}
\usepackage[ruled, vlined]{algorithm2e}
\SetAlgoCaptionLayout{justify}
\usepackage{hhline}
\usepackage{quantikz}
\usepackage{wrapfig}
\usepackage[
	raiselinks=true,                 
	bookmarks=true,                     
	bookmarksopenlevel=1,               
	bookmarksopen=true,
	bookmarksnumbered=true,
	hyperindex=true,
	plainpages=false,
	pdfpagelabels=true,
	pdfborder={0 0 0.5},
	colorlinks=true,
	linkcolor={teal},
	citecolor={teal},
	urlcolor={teal},
]{hyperref}
\usepackage{footmisc}
\usepackage[capitalise]{cleveref}
\crefformat{footnote}{#2\footnotemark[#1]#3}

\usepackage[nolist, nohyperlinks]{acronym}
\newacro{qft}[QFT]{quantum Fourier transform}
\newacro{fem}[FEM]{finite element method}

\begin{document}

\title{Quantum oracles for the finite element method} 

\author{Sven Danz}
\email{s.danz@fz-juelich.de}
\affiliation{Institute for Quantum Computing Analytics (PGI-12), Forschungszentrum J\"ulich, 52425 J\"ulich, Germany}
\affiliation{Theoretical Physics, Saarland University, 66123 Saarbrücken, Germany}

\author{Tobias Stollenwerk}
\affiliation{Institute for Quantum Computing Analytics (PGI-12), Forschungszentrum J\"ulich, 52425 J\"ulich, Germany}

\author{Alessandro Ciani}
\affiliation{Institute for Quantum Computing Analytics (PGI-12), Forschungszentrum J\"ulich, 52425 J\"ulich, Germany}

\begin{abstract}
In order to assess potential advantages of quantum algorithms that require quantum oracles as subroutines, the careful evaluation of the overall complexity of the oracles themselves is crucial.
	This study examines the quantum routines required for the implementation of oracles used in the block-encoding of the $N \times N$ stiffness and mass matrices, which typically emerge in the finite element analysis of elastic structures.
	Starting from basic quantum adders, we show how to construct the necessary oracles, which require the calculation of polynomials, square root and the implementation of conditional operations.
	We propose quantum subroutines based on fixed-point arithmetic that, given an $r$-qubit register, construct the oracle using $\bigO((K + L + N_{\mathrm{geo}} + N_{\mathrm{D}}) r)$ ancilla qubits and have a $\bigO((K + L)r^2 + \mathrm{log}_2(N_{\mathrm{geo}} + N_{\mathrm{D}}))$ runtime, with $K$ the order at which we truncate the polynomials, $L$ the number of iterations in the Newton-Raphson subroutine for the square root, while $N_{\mathrm{geo}}$ and $N_{\mathrm{D}}$ are the number of hypercuboids used to approximate the geometry and the boundary, respectively.
	Since in practice $r$ scales as $r = \mathcal{O}(\log_2 N)$, and assuming that the other parameters are fixed independently of $N$, this shows that the oracles, while still costly in practice, do not endanger potential polynomial or exponential advantages in $N$.
\end{abstract}
\maketitle

\section{Introduction}
\label{sec:intro}
Many promising quantum algorithms that provide a quantum advantage over classical methods assume the efficient implementation of a quantum oracle~\footnote{
	Oracles are black-box quantum operations that are assumed to be implementable without providing an explicit gate construction.
}.
In fact, the runtimes of these algorithms are often quantified in terms of the number of oracle queries, which establishes the \emph{query complexity} of the algorithm.
In contrast, the \emph{gate complexity} of an algorithm quantifies the runtime in terms of the number of elementary gates either quantum or classical.
The implicit assumption behind the query model is that an oracle call has unit cost independently of whether the oracle is classical or quantum.
Therefore, whether a potential quantum advantage in the query model carries over to the gate model is contingent upon the efficient implementation of such quantum oracles in terms of elementary quantum gates.
Whether this is the case or not depends on the particular use case at hand.
Thus, it is crucial to identify the use cases for which an efficient implementation of the quantum oracles is possible in order for quantum algorithms to be successful in real world applications.

The use of oracles in quantum computing can be traced back to the
Deutsch-Jozsa algorithm~\cite{deutsch1985,deutsch1992}, in which the oracle computes a potentially balanced function.
Grover's algorithm~\cite{grover1996} for database search also assumes access to an oracle that marks computational basis states of interest.
In fact, Grover's algorithm achieves a quadratic quantum advantage in the query sense compared to classical algorithms.
In recent years, several quantum algorithms have been re-formulated in the unified framework of quantum signal processing (QSP) and quantum singular-value transformation (QSVT)~\cite{martyn2021, gilyen2019, dalzell2023, lowChuang2017, Low2019hamiltonian}.
The fundamental primer of this framework is the concept of block-encoding of a generic matrix $\bm{H}$ into a block of a (larger) unitary $U$.
One way to construct such block-encodings requires the availability of matrix oracles that encode information about the position and the value of the non-zero matrix elements of $\bm{H}$ (see Refs.~\cite{childs2010,berry2012,gilyen2019, Low2019hamiltonian, dalzell2023} for examples).
This constitutes the so-called sparse-access model to the matrix information, which assumes that the matrix has bounded sparsity $s$ and each matrix element can be computed efficiently with a reversible quantum circuit.
In principle, it is always possible to achieve reversible versions of every oracle that can be implemented classically.
This is because we can implement the NAND (NOT-AND) gate, a universal gate for classical computing, using a Toffoli and a NOT gate.
However, this approach requires many ancilla qubits potentially endangering the memory footprint.
Furthermore, aiming for quantum algorithms with speed-up over their classical equivalents, tightens the definition of \emph{efficient}, meaning, an oracle that is efficient for the use in a classical algorithm could be a bottleneck in a faster quantum algorithm.

In this work, we focus on the efficient construction of matrix oracles that appear in the \ac{fem}~\cite{fish2007, Ern2004, bathe2006, cook1989,rao2005} for solving partial differential equations.
\ac{fem} simulations are ubiquitous in science and engineering and routinely used at the industrial level for fluid dynamics, electromagnetic and structural simulations, just to name a few examples.
Thus, while it is natural to look at possible applications of quantum computing to \ac{fem} simulations~\cite{clader2013, montanaro2016}, it is first necessary to show how \ac{fem} simulations can be efficiently embedded in a quantum computer in the oracle model.
For concreteness, we focus on the construction of the oracles in the context of normal mode analysis of solid structures.
The application of quantum computing to this problem has been recently studied in Ref.~\cite{danz2024calculating}, where however, it was analyzed already assuming a discretized form as a system of coupled harmonic oscillators.
Once discretized, the dynamic behavior of the problem is completely defined by the so-called mass $\bm{M}$ and stiffness $\bm{K}$ matrices
\begin{equation}
	\frac{\diff^2 }{\diff t^2}\vectorarrow{z}
	=
	-\underbrace{\bm{M}^{-1/2}\bm{K}\bm{M}^{-1/2}}_{\equiv \bm{H}} \vectorarrow{z}.
	\label{eq:generalized_epp}
\end{equation}
This problem can be mapped to a quantum computer using two oracles for the sparse matrix $\bm{H}$.
The first is for the determination of the non-zero elements in a given row $u$ of $\bm{H}$.
In structural problems, the non-zero positions are not random, but follow strict rules.
Assuming that we have a system with high symmetry (a regular mesh), the number of rules, necessary to describe all those positions, stays minimal.
Hence, a small superposition, generated by a Walsh-Hadamard gate, followed by an $u$-dependent shift (addition) is usually sufficient.

The second oracle $O_\vartheta$ computes the binary sign $\sgnb{a\geq 0}=0, \sgnb{a<0}=1$ of $H_{uv}$ and a related angle $\vartheta_{uv}$ defined by
\begin{equation}
	\vartheta_{uv}
	=
	\arccos\sqrt{\frac{
			\abs{H_{uv}}
		}{
			\norm{H}_\mathrm{max}
		}},
	\label{eq:angle}
\end{equation}
with $\norm{H}_\mathrm{max}=\max_{uv}\abs{H_{uv}}$, and stores them in $r+1$ qubits
\begin{equation}
	O_\vartheta\ket{u,v}\ket{0}^{\otimes (r+1)}
	=
	\ket{u,v}\ket{\sgnb{H_{uv}},\vartheta_{uv}}.
	\label{eq:oracle}
\end{equation}
We further simplify the arccosine in \cref{eq:angle}.
A Taylor expansion transforms it into a polynomial in $\sqrt{x}$ function\footnote{
	Note that this is formally a signomial function, which is similar to polynomial functions but allow non-integer exponents for positive variables.
}
\begin{equation}
	\arccos{\sqrt{x}}
	=
	\frac{\pi}{2}
	-\sqrt{x}
	-\frac{3}{40}\sqrt{x}^3
	-\frac{5}{112}\sqrt{x}^5
	+\dots,
	\label{eq:arccos}
\end{equation}
that we truncate at order $K$ terms assuming that all $x$ of interest satisfy $x\ll 1$.
This method is not accurate for $x\approx 1$.
However, it is sometimes possible to renormalize $x$ in trade for a lower accuracy or one can replace $\arccos \sqrt{x}$ with $2\arcsin\sqrt[4]{(1-x)/2}$, which is equal, but its arguments are limited to $[0,1/\sqrt[4]{2}]$ (cf. Ref.~\cite{Haner2018}).
The Taylor expansion of the arcsine is similar to the arccosine which is why we stick with the requirements of \cref{eq:arccos} for simplicity.

The purpose of this manuscript is to demonstrate how the quantum computation of $\vartheta_{uv}$ scales with respect to the matrix size $N$, and other relevant parameters, in typical scenarios.
Additionally, the manuscript aims to provide a comprehensive account of the quantum routines that are necessary to perform these computations and their associated computational complexity.
A previous analysis of the computations, necessary for implementation of related quantum oracles, was done, with fewer details, in Ref.~\cite{Haner2018}.
Also, we highlight the work of Ref.~\cite{camps2024}, which, in the same spirit of our work, provides an explicit construction for the block-encoding of certain structured sparse matrices, but without a connection to \ac{fem}.

\subsection{Structure}
The paper is structured as follows.
In \cref{sec:compl}, we discuss the existing literature describing arithmetic operations.
Furthermore, we extend it by a detailed description of how to compute polynomial and square root (and inverse square root) functions based on quantum adders, which is required by \cref{eq:angle,eq:arccos} assuming that $H_{uv}$ can be approximated via the use of polynomials and square root as well.
This contains a complete discussion of the corresponding computational complexities.
In \cref{sec:cont_prob}, we go one step back, and derive $\bm{M}$ and $\bm{K}$ using the \ac{fem} before simplifying them.
This allows for an efficient computation of $H_{uv}$, based on the arithmetic operations introduced in \cref{sec:compl}.
\cref{sec:application} describes the implementation of $O_\vartheta$ for a homogeneous one-dimensional example and its computational complexity.
We conclude this manuscript with a discussion of our results in \cref{sec:conclusion}.

In \cref{sec:exponentiation} we give a more general method for the computation of signomial functions that relies on exponentiation and \emph{in-place} arithmetic.
The latter is described in \cref{sec:in-place} for the examples of a square operation and a multiplication, which allows us to reduce the memory requirements.
The computation of $H_{uv}$ requires logic operations given in \cref{sec:logic_operator}.
At last in \cref{sec:adv_contours}, we give an outlook on how to construct more complicated geometries than those discussed in \cref{sec:application}.

\section{Quantum routines and their complexity}
\label{sec:compl}

In this section, we propose quantum routines for the computation of the functions (cf. \cref{eq:arccos}), that are required in oracles used for block-encoding of \ac{fem} based matrices.
Classically one would compute the series in \cref{eq:arccos} in two steps: the computation of the square root followed by the computation of a polynomial.
The quantum arithmetic operations, necessary for those two routines, and their complexity are described in detail in this section.
In order to preserve potential polynomial or exponential advantages in algorithms, that rely on this oracle, it is necessary that the oracle scales polylogarithmically in $N$~\cite{danz2024calculating}.
We begin by reviewing state-of-the-art methods in \cref{sec:arithmetic}.
Both, the computation of polynomials and the square root, require the primitives addition and multiplication.
We construct both these routines from quantum adders.
This is described generically in \cref{sec:polynomials,sec:newton-raphson} respectively.

\subsection{Adders}
\label{sec:arithmetic}
\begin{figure}
	\centering
	\includegraphics[width = 40mm]{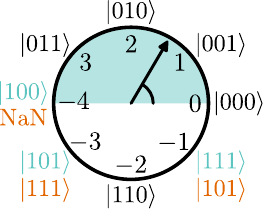}
	\caption{
		Comparison of the two's complement (\blue{blue}, top string) and the sign-magnitude representation (\orange{orange}, bottom string).
		Both forms share the same representation for positive numbers but are inverted for negative numbers.
		Here, $r=2$ and $p=0$.
	}
	\label{fig:forms}
\end{figure}

In this manuscript, we focus on the so-called fixed-point arithmetic encoded in the computational basis.
To be more precise, we use the two's complement form most of the time to encode numbers $b\in[-2^{r-p},2^{r-p}-2^{-p}]$ in $r+1$ qubits.
$p\in [0, \dots, r]$ is the binary point position in relation to the least significant qubit.
The mapping between $b$ and its two's complement form is summarized by the following mapping:
\begin{multline}
	2^p b \mod 2^{r+1}
	=
	\left(
	\sum_{k=0}^{r} 2^{k}b_{k+1}
	\right)
	+ 2^p\varepsilon_b
	\\
	\to \ket{b_{r+1},\dots,b_1}
	\equiv
	\ket{b},
	\label{eq:bin_form}
\end{multline}
where $\{b_k\in\{0,1\},\forall k\in [r+1]\}$\footnote{
	Here, $[r]=\{1,\dots, r\}$ is the set of the lowest $r$ natural numbers.
	\label{fn:natural}
}
are the binary digits of $b$.
This representation comes with an error $\abs{\varepsilon_b}\leq 2^{-(p+1)}$.
In what follows, $r+1$ will be the default size of all registers used to store numbers.
We further assume, that $r$ and $p$ are fixed and satisfy $\abs{b}\ll 2^{r-p}$ for all numbers $b$.
This prevents overflow due to addition, multiplication or more advanced computations.
The two's complement form is mostly used in the following because it handles naturally the addition of negative numbers.
However, another standard representation is the sign-magnitude form, that is beneficial for multiplications.
We compare the two representations in \cref{fig:forms}.

The basis of every arithmetic operation is the addition.
The corresponding in-place\footnote{
	The alternative are \emph{out-of-place} operations which require an additional register for the result.
} quantum routine $\mathrm{ADD}$ would read two quantum register with numbers encoded (e.g. $a$ and $b$) and replace one of them with the sum of both
\begin{equation}
	\mathrm{ADD}\ket{a}\ket{b}
	=
	\ket{a}\ket{a+b}.
\end{equation}
The register with altered value is called target register.

\begin{figure}
	\centering
	\begin{adjustbox}{max width=\linewidth}
		\begin{quantikz}[wire types={b,b,b},classical gap=0.07cm]
			\lstick{$|a\rangle$} &\gate[3,label style={yshift=5mm}]{\mathrm{ADD}}&\rstick{$|a\rangle$}
			\\
			\lstick{$\blue{|0\rangle}$} &&\rstick{$\blue{|0\rangle}$}
			\\
			\lstick{$|b\rangle$} &\gateoutput{$\rightarrow$}&\rstick{$|a+ b\rangle$}
		\end{quantikz}
	\end{adjustbox}
	\caption{
		Generic in-place quantum adder.
		The $\rightarrow$ symbol marks the target register.
		The center register in \blue{blue} is needed to take care of the carries and is just large enough to satisfy the ancilla qubit requirements of the used routine (cf. \cref{tab:adder}).
	}
	\label{fig:modadd}
\end{figure}
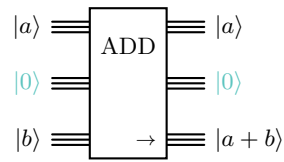

It is important to chose $r$ and $p$ carefully to prevent overflow (i.e. $a+b<2^{r-p}$ or more general $a,b \ll 2^{r-p}$).
The gate symbol for the in-place adder is shown in \cref{fig:modadd} in which we added a possible ancilla register that may be necessary for some routines.

Based on the basic adder one can find modified versions.
Every binary quantum adder can be easily transformed into a subtractor SUB by running it backwards (i.e. $\mathrm{SUB}=\mathrm{ADD}^\dagger$).
Furthermore, one can replace the quantum register $\ket{a}$ with a classical register achieving a \emph{quantum-classical} adder $\mathrm{qcADD}_l$ (subtractor $\mathrm{qcSUB}_l$), that adds (subtracts) the fixed number $l$ encoded in the classical register
\begin{subequations}
	\begin{align}
		\mathrm{qcADD}_l\ket{b}
		 & =
		\ket{b+l},
		\\
		\mathrm{qcSUB}_l\ket{b}
		 & =
		\ket{b-l}.
	\end{align}
\end{subequations}
Knowing the number $l$ classically allows us to replace controlled operations connecting the classical register $\ket{l}_c$ with the quantum register $\ket{b}$ and (or) the ancilla registers with operations that act solely on the quantum registers.
For instance a CNOT with control a bit in the classical register and target one in a quantum register would be replaced by the identity if the classical bit is $0$ and by a NOT gate if it is $1$.
This makes quantum-classical routines faster compared to their \emph{quantum-quantum} counterparts.

For the purpose of simple addition, one can use different algorithms.
Classical methods that can be also implemented on quantum devices are already known in the literature.
One of those methods is the carry-ripple addition, a rather slow but simple and memory efficient method that adds two numbers bit-by-bit.
Two quantum versions of this algorithm are described in Refs.~\cite{vedral1996,beckman1996}.
The runtime of those algorithms scale linearly with the number of binary digits $r$ (no separate sign qubit), but require one additional quantum register of size $\bigO(r)$ for the carry.
\begin{table}
	\centering
	\caption{
		Quantum adder routines for $r$-digit summands.
		The runtimes and memory requirements are shown for representative specific implementations of each type of adder (see discussion in the main text).
	}
	\label{tab:adder}
	\begin{tabular}{lcc}
		\hline
		Method                                  & Runtime $t_\mathrm{ADD}$ & Memory $n_\mathrm{ADD}$
		\\
		                                        &                          & in qubits
		\\
		\hline
		Carry-ripple~\cite{cuccaro2004}         & $\bigO(r)$               & $2r+1$
		\\
		Carry-select~\cite{thomsen2008}         & $\bigO(\sqrt{r})$        & $2r+\sqrt{r}$
		\\
		Conditional sum~\cite{vanmeter2005}     & $\bigO(\log_2{r})$       & $\approx 7r-6$
		\\
		Carry-lookahead~\cite{draper2004}       & $\bigO(\log_2{r})$       & $ 4r-\log_2 r-1 $
		\\
		Fourier transform~\cite{Ruiz-Perez2017} & $\bigO(r^2)$             & $2r$
		\\
		\hline
	\end{tabular}
\end{table}
A more advanced version of this method is described in Ref.~\cite{cuccaro2004}, that requires only one additional ancilla qubits on top of the $2r$ qubits for the two summands (cf. \cref{tab:adder}).
Finally, Ref.~\cite{gidney2018} halved the runtime cost compared to Ref.~\cite{cuccaro2004} at the price of re-introducing $r$ ancillas.

One disadvantage of the carry-ripple method is that most of the qubits are idle most of the time.
This can be prevented by parallelization methods also known from classical computing.
Those reduce the total runtime while increasing the memory requirements.
The first of those is the carry-select adder, that reduces the total runtime to $\bigO(\sqrt{r})$.
It does so by splitting the summands bitwise into $p$ packs of $q$ digits.
The partial $q$-qubit adders are executed for both possible carry input values (0 and 1).
This allows us to combine them later with $p$ so-called multiplexer that choose the correct result depending on the carry from the previous package.
The total runtime scales like $\bigO(p+q)$, which is minimal for $p=q=\sqrt{r}$.
This method was first proposed in Ref.~\cite{zalka1998} and later fully described in different versions in Refs.~\cite{vanmeter2005,thomsen2008}.

The carry-select method can be further improved by using a cascade-like instead of a linear multiplexer structure.
This allows us to pre-combine number packages parallel before they get combined again.
In this way, the runtime can be reduced to $\bigO(q+\log_2{p})$ which has optimal scaling $\bigO(\log_2{r})$ for $q=1$ and $p=r$.
That method is called conditional sum addition in Ref.~\cite{vanmeter2005}.

Another classical method is the carry-lookahead addition.
The idea is to split the computation of the carry and the sum into two parts executing one after the other.
This yields knowledge of all carries before the bitwise addition starts, which allows for a parallel execution of them dropping their runtime to $\bigO(1)$.
The runtime of the computation of the carries scales like $\bigO(\log_2 r)$ due to a cascade-like structure.
This method was first proposed for quantum devices in Ref.~\cite{draper2004} and later further analyzed in Ref.~\cite{vanmeter2005}.

A completely different approach, that makes use of the unique properties of quantum states, is the \ac{qft} adder originally proposed in Ref.~\cite{draper2000} and later extended in Ref.~\cite{Ruiz-Perez2017}.
This method uses the \ac{qft} to move one of the two summands into the phase of the quantum state, which can be increased or decreased by phase rotations.
If controlled by the other summand, this yield an intuitive algorithm for addition and subtraction.
An application of the inverse \ac{qft} at the end moves the sum back into the base of the quantum state.
Both, the \ac{qft} and the controlled phase rotations have $\bigO(r^2)$ runtimes, which is slower than all other methods discussed in this manuscript.
However, it achieves this by using the minimum of only $2r$ qubits.
The \ac{qft} adder can be mapped onto a Toffoli based adder reducing the runtime to $\bigO(r)$~\cite{Paler2022}.
The runtimes and memory requirements for all those methods are gathered in \cref{tab:adder}.

\begin{figure}
	\subfloat[Multiple addition ($\mathrm{ADD}^{2^{k}}$)]{
	\label{fig:modadd_k_a}
	\centering
	\begin{adjustbox}{max width=\linewidth}
		\begin{quantikz}[wire types={b,b,b,b,b},classical gap=0.07cm]
			\lstick{$|a_{r},\dots,a_{r-k+1} \rangle =|0\rangle^{\otimes k}$} &\qwbundle{k}&\gategroup[
				5,
				steps=1,
				style={
						dashed,
						rounded corners,
						fill=teal!30,
						inner xsep=0pt
					},
				background,
				label style={
						label position=above,
						anchor=south,
						yshift=-0.15cm,
						align=center
					}
			]{$\mathrm{ADD}^{2^k}$}&\rstick[2]{$|a \rangle$}
			\\
			\lstick{$|a_{r-k},\dots,a_{1}\rangle$}&\qwbundle{r-k} &\gate[3,label style={yshift=4mm}]{\mathrm{ADD}}&
			\\
			\lstick{$\blue{|0\rangle}$} &&&\rstick{$\blue{|0\rangle}$}
			\\
			\lstick{$|b_{r},\dots,b_{k+1}\rangle$}&\qwbundle{r-k} &\gateoutput{$\rightarrow$}&\rstick[2]{$|2^ka+ b\rangle$}
			\\
			\lstick{$|b_{k},\dots,b_{1} \rangle$}&\qwbundle{k} &&
		\end{quantikz}
	\end{adjustbox}
	}

	\subfloat[Fractional addition ($\mathrm{ADD}^{2^{-k}}$)]{
	\label{fig:modadd_k_b}
	\centering
	\begin{adjustbox}{max width=0.95\linewidth}
		\begin{quantikz}[wire types={b,b,b,b},classical gap=0.07cm]
			\lstick{$|a_{r},\dots,a_{k+1} \rangle$} &\qwbundle{r-k} &\gate[2,swap]{}\gategroup[
			4,
			steps=3,
			style={
					dashed,
					rounded corners,
					fill=teal!30,
					inner xsep=0pt
				},
			background,
			label style={
					label position=above,
					anchor=south,
					yshift=-0.15cm,
					align=center
				}
			]{$\mathrm{ADD}^{2^{-k}}$}&\qwbundle{k}&\gate[2,swap]{}&\rstick[2]{$|a \rangle$}
			\\
			\lstick{$|a_{k},\dots,a_{1}\rangle$}&\qwbundle{k} &&\gate[3,label style={yshift=4mm}]{\mathrm{ADD}}&&
			\\
			\lstick{$\blue{|0\rangle}$} &&&&&\rstick{$\blue{|0\rangle}$}
			\\
			\lstick{$|b\rangle$}&\qwbundle{r} &&\gateoutput{$\rightarrow$}&&\rstick{$|2^{-k}a+ b\rangle$}
		\end{quantikz}
	\end{adjustbox}
	}
	\caption{
		Modified versions of a quantum adder ($\mathrm{ADD}^{2^k}$), that allow us to add multiple of $a$ if the multiplier is a power of $2$.
		For $k>0$ (a), the method requires the $k$ most significant qubits of $\ket{a}$ to be $\ket{0}$ as it neglects them.
		The remaining $r-k$ digits are added to the $r-k$ most significant digits of $b$.
		For $k<0$ (b), it adds the $r-k$ most significant bits to $\ket{b}$.
	}
	\label{fig:modadd_k}
\end{figure}
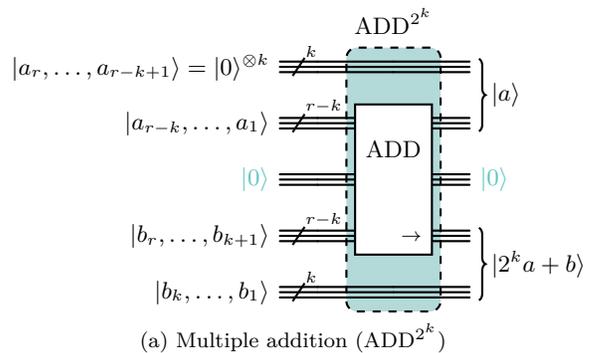
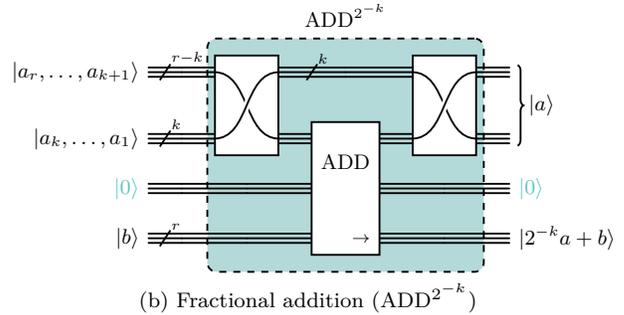

Furthermore, starting from the basic in-place adder ADD in \cref{fig:modadd}, we can construct a modified version that adds a multiple or a fraction of one of the two summands to the other as long as the factor is a power of $2$:
\begin{equation}
	\mathrm{ADD}^{2^{\pm k}}\ket{a}\ket{b}
	=
	\ket{a}\ket{2^{\pm k}a+b}.
\end{equation}
For this we shift the significance of the bits of the two numbers before adding them.
In the case of the addition of a multiple of $a$, this means we reduce ADD to an $r-k$ qubit version and apply it to the $r-k$ least significant qubits of the first register and to the $r-k$ most significant qubits of the second register (see \cref{fig:modadd_k_a}).
This process ignores the first $k$ digits of $a$.
Hence, those digits should be zero to prevent information loss.
This matches our overflow condition for positive numbers, requiring $\abs{a}\ll 2^{r-p}$.
For the addition of the fraction $2^{-k}a$ we add the $r-k$ most significant bits of $\ket{a}$ to $\ket{b}$.
Here, the $k$ least-significant digits of $a$ can be non-zero as their contribution vanishes due to the finite precision with $r$ qubits.
Both methods result in a shift of $k$ digits between the two binary numbers and with it either in the sum $2^ka+b$ or $2^{-k}a+b$.
The additional factor will come in handy for the multiplication.
This routine works for arbitrary $p$ as long as $a\geq 0$.

\subsection{From adders to polynomials}
\label{sec:polynomials}

We mentioned before that addition can be taken as the basis of every other arithmetic operation.
In this section, we show this by constructing a multiplier from a generic adder, and compute a polynomial function (cf.~\eqref{eq:arccos}) from all other primitives.
The described method is just one way of doing this and shall give a general understanding of how one can achieve all these operations starting with an arbitrary adder.

\begin{figure*}
	\centering
	\begin{adjustbox}{max width=\linewidth}
		\begin{quantikz}[wire types={q,q,q,q,q,b,b,b,q},classical gap=0.07cm]
			\lstick{$|a_{r+1}\rangle$} &\ctrl{1}\gategroup[
				4,
				steps=2,
				style={
						dashed,
						rounded corners,
						fill=teal!10,
						inner xsep=2pt
					},
				background,
				label style={
						label position=above,
						anchor=south,
						yshift=-0.15cm,
						align=center
					}
			]{two's complement to\\sign-magnitude form}&\ctrl{3}&\ctrl{8}\gategroup[
				9,
				steps=2,
				style={
						dashed,
						rounded corners,
						fill=teal!25,
						inner xsep=2pt
					},
				background,
				label style={
						label position=below,
						anchor=north,
						yshift=-0.15cm
					}
			]{$\sgn{ab}$}&&&\ \ldots\ &&\ctrl{3}\gategroup[
				4,
				steps=2,
				style={
						dashed,
						rounded corners,
						fill=teal!10,
						inner xsep=2pt
					},
				background,
				label style={
						label position=above,
						anchor=south,
						yshift=-0.15cm,
						align=center
					}
			]{sign-magnitude to\\two's complement form}&\ctrl{1}&\rstick[4]{$|a\rangle$}
			\\
			\lstick{$|a_{r}\rangle$} &\gate[3,label style={yshift=4mm}]{\mathrm{qcSUB}_1}&\targ{}&&&\gategroup[
				7,
				steps=3,
				style={
						dashed,
						rounded corners,
						fill=teal!40,
						inner xsep=2pt
					},
				background,
				label style={
						label position=below,
						anchor=north,
						yshift=-0.65cm
					}
			]{absolute product $\abs{a}\abs{b}$}&\ \ldots\ &\ctrl{4}&\targ{}&\gate[3,label style={yshift=4mm}]{\mathrm{qcADD}_1}&
			\\
			\wave &&&&&&&&&&&
			\\
			\lstick{$|a_1\rangle$} &&\targ{}&&&\ctrl{2}&\ \ldots\ &&\targ{}&&
			\\
			\lstick{$|b_{r+1}\rangle$} &\ctrl{1}\gategroup[
				2,
				steps=2,
				style={
						dashed,
						rounded corners,
						fill=teal!10,
						inner xsep=2pt
					},
				background,
				label style={
						label position=above,
						anchor=south,
						yshift=-0.15cm
					}
			]{}&\ctrl{1}&&\ctrl{4}&&\ \ldots\ &&\ctrl{1}\gategroup[
				2,
				steps=2,
				style={
						dashed,
						rounded corners,
						fill=teal!10,
						inner xsep=2pt
					},
				background,
				label style={
						label position=above,
						anchor=south,
						yshift=-0.15cm
					}
			]{}&\ctrl{1}&\rstick[2]{$|b\rangle$}
			\\
			\lstick{$|b_{r},\dots,b_1\rangle$} &\gate{\mathrm{qcSUB}_1}&\targ{}&&&\gate[3,label style={yshift=4mm}]{\mathrm{ADD}^{2^{-p}}}&\ \ldots\ &\gate[3,label style={yshift=4mm}]{\mathrm{ADD}^{2^{r-p-1}}}&\targ{}&\gate{\mathrm{qcADD}_1}&
			\\
			\lstick{$\blue{|0\rangle}$} &&&&&&\ \ldots\ &&&&\rstick{$\blue{|0\rangle}$}
			\\
			\lstick{$|0\rangle^{\otimes r}$} &&&&&\gateoutput{$\rightarrow$}&\ \ldots\ &\gateoutput{$\rightarrow$}&\targ{}\gategroup[
				2,
				steps=2,
				style={
						dashed,
						rounded corners,
						fill=teal!10,
						inner xsep=2pt
					},
				background,
				label style={
						label position=above,
						anchor=south,
						yshift=-0.15cm
					}
			]{}&\gate{\mathrm{qcADD}_1}&\rstick[2]{$|ab\rangle$}
			\\
			\lstick{$|0\rangle$} &&&\targ{}&\targ{}&&\ \ldots\ &&\ctrl{-1}&\ctrl{-1}&
		\end{quantikz}
	\end{adjustbox}
	\caption{
		Quantum multiplier (MUL) based on the modified quantum adder ($\mathrm{ADD}^{2^k}$) from \cref{fig:modadd_k}.
		It starts with a transformation from the two's complement into the sign-magnitude form.
		Next we compute the relative sign, followed by the product of the magnitudes.
		The ancilla register in \blue{blue} is just large enough to satisfy the ancilla requirements of the adders.
		Here, we neglect that the adders and subtractors in the form transformations may also require ancilla qubits.
	}
	\label{fig:circ_mult}
\end{figure*}
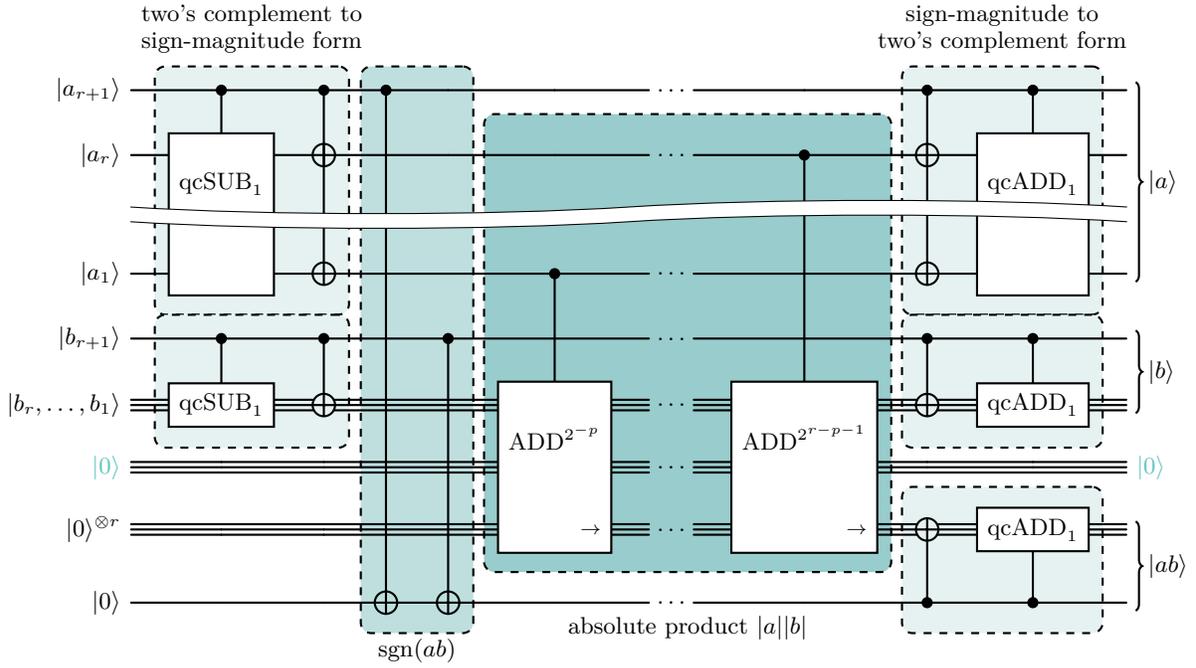

\paragraph{Multiplier}
The next routine we consider is the quantum multiplier MUL, which adds the product of two numbers (e.g. $a$ and $b$) to a target register
\begin{equation}
	\mathrm{MUL}\ket{a,b}\ket{z}
	=
	\ket{a,b}\ket{z+ab}.
\end{equation}
Any product can be deconstructed into a relative sign and a sum requiring only addition
\begin{equation}
	ab
	=
	\sgn{ab}
	\sum_{k=0}^{r-1}
	2^{k-p}\abs{a}_{k+1}\abs{b}.
	\label{eq:product}
\end{equation}
Here, $\abs{a}_{k+1}$ represents the digits of the fixed-point representation of $\abs{a}$, which
control if the summand $2^{k-p}\abs{b}$ contributes to the sum or not.
Hence, we compute the product in three steps (see \cref{fig:circ_mult}).
First, we convert $\ket{a}$ (and $\ket{b}$) from their two's complement form into the sign-magnitude form $\ket{\sgnb{a},\abs{a}_r,\dots,\abs{a}_1}$.
This allows us to access the sign and the magnitude directly.
The transformation consists of two parts, a subtraction $\mathrm{qcSUB}_1$ and a $\mathrm{NOT}^{\otimes r}$ gate, both controlled by $\ket{a_{r+1}}$ and applied to $\ket{a_r,\dots,a_1}$.
We provide a simple example for clarity ($r=2$, cf. \cref{fig:forms})
\begin{equation}
	b
	=
	-3
	\to
	\ket{101}
	\xrightarrow{\mathrm{qcSUB}_1} \ket{100}
	\xrightarrow{\mathrm{NOT}^{\otimes 2}} \ket{111}.
\end{equation}
A broader discussion of converters between those two forms is given in Ref.~\cite{Orts2019}.
After the transformation, the calculation of the relative sign $\sgn{ab}$ is a matter of two NOT gates applied the most significant bit of the target register and controlled by $\ket{a_{r+1}}$ and $\ket{b_{r+1}}$ (cf. \cref{fig:circ_mult}).
Finally, it remains to compute the sum over $2^{k-p}\abs{b}$ controlled by the single qubits $\ket{\abs{a}_k}$, which equals the product of the absolutes.

The sum is stored in the $r$ least-significant bits of the target register.
In the end, we need to re-transform all register into the two's complement form.
It is important that we choose $r$ and $p$ appropriate to store any product $ab$, and later also $\sum_k c_kb^k$ for the polynomial, without overflow.

A qubit-shift-based modular multiplication of two positive integers, stored in qubits, was already described in Ref.~\cite{rieffel2014}.
We further extended the algorithm for negative fixed point numbers.

The multiplication requires one additional register of size $r+1$ giving a total of $n_{\mathrm{MUL}}$ qubits, which containing
$n_{\mathrm{MUL,a}}$ ancillas
\begin{subequations}
	\begin{align}
		n_{\mathrm{MUL}}
		 & =
		3(r+1)+n_{\mathrm{ADD,a}},
		\\
		n_{\mathrm{MUL,a}}
		 & =
		n_{\mathrm{ADD,a}},
	\end{align}
\end{subequations}
Here, $n_{\mathrm{ADD,a}}$ is the adder specific ancilla requirement.
The runtime of this routine is
\begin{multline}
	t_{\mathrm{MUL}}
	=
	2t_\mathrm{qcADD}
	+(2r+2)t_\mathrm{CNOT}
	+\sum_{k=0}^{r-1} t_{\mathrm{ADD}^{2^{k-p}}}
	\\
	=
	\bigO(rt_{\mathrm{ADD}}),
	\label{eq:lat_mult}
\end{multline}
where $t_{\mathrm{ADD}}$ is the runtime of the chosen $r$ qubit adder (cf. \cref{tab:adder}).
$t_\mathrm{CNOT}$, $t_\mathrm{qcADD}$, and $t_{\mathrm{ADD}^{2^{k-p}}}$ are the runtimes of one CNOT gate, $\mathrm{qcADD}_1$, and the modified adder $\mathrm{ADD}^{2^{\pm k}}$ respectively.
Here, we use that the adder dominate the runtime and all modified adders require less runtime than the basic ADD.

\paragraph{Horner's scheme based polynomial}
With a working routine for the multiplication of negative fixed-point numbers we can go on and compute polynomials.
Here we will follow Horner's scheme, similar to \cite{Haner2018}, which decomposes polynomials efficiently by minimizing the total number of multiplications.
\begin{multline}
	\mathrm{poly}(b)
	=
	\sum_{k=0}^{K}c_k b^k
	\\
	=
	c_0+b\left(c_1+b\left(c_2+\dots + b\left(c_{K-1}+bc_K\right)\dots\right)\right).
	\label{eq:Horner}
\end{multline}
This routine becomes ineffective for polynomials with only few non-zero coefficients $c_k$.
However, for the polynomial at hand \cref{eq:arccos} only half of all coefficients are zero.

\begin{figure*}
	\centering
	\begin{adjustbox}{max width=\linewidth}
		\begin{quantikz}[wire types={b,b,b,b,b},classical gap=0.07cm]
			\lstick{$|b\rangle$}&\gategroup[
				5,
				steps=7,
				style={
						dashed,
						rounded corners,
						fill=teal!30,
						inner xsep=2pt
					},
				background,
				label style={
						label position=above,
						anchor=south,
						yshift=-0.15cm
					}
			]{$U_{K\to1}$}&\gate[2]{\mathrm{qcMUL}(c_K)}&\gate[3]{\mathrm{MUL}}&\gate[2, swap]{}&&\ \ldots\ &&&&\gate[5]{U_{K\to 1}^\dagger}&\rstick{$|b\rangle$}
			\\
			\lstick{\blue{$|0\rangle^{\otimes (r+1)}$}} &\gate{\prod_{i=1}^{r+1}X_i^{c_{K-1,i}}}&\gateoutput{$\rightarrow$}&&&\gate[3]{\mathrm{MUL}}&\ \ldots\ &&&&&\rstick{\blue{$|0\rangle^{\otimes (r+1)}$}}
			\\
			\lstick{\blue{$|0\rangle^{\otimes (r+1)}$}} &\gate{\prod_{i=1}^{r+1}X_i^{c_{K-2,i}}}&&\gateoutput{$\rightarrow$}&&&\ \ldots\ &&&&&\rstick{\blue{$|0\rangle^{\otimes (r+1)}$}}
			\\
			\wave &&&&&&&\gate[2]{\mathrm{MUL}}&\gate[3]{\mathrm{MUL}}&&&&&
			\\
			\lstick{\blue{$|0\rangle^{\otimes (r+1)}$}} & \gate{\prod_{i=1}^{r+1}X_i^{c_{1,i}}}&&&&&\ \ldots\ &\gateoutput{$\rightarrow$}&&&&\rstick{\blue{$|0\rangle^{\otimes (r+1)}$}}
			\\
			\lstick{$|0\rangle^{\otimes (r+1)}$} &\gate{\prod_{i=1}^{r+1}X_i^{c_{0,i}}}&&&&&\ \ldots\ &&\gateoutput{$\rightarrow$}&&&\rstick{$|\mathrm{poly}(b)\rangle$}
		\end{quantikz}
	\end{adjustbox}
	\caption{
		Quantum operation for the computation of polynomials (POLY) based on Horner's scheme.
		This requires both a quantum-quantum multiplier (MUL) and a quantum-classical multiplier (qcMUL).
		The ancilla qubits used by them are hidden here for illustrative reasons.
		The last gate represents the Hermitian conjugate $U_{K\to1}^\dagger$ of all gates in the \blue{blue} box.
		It reinitializes the ancilla registers and returns the input register to $\ket{b}$.
	}
	\label{fig:.poly}
\end{figure*}
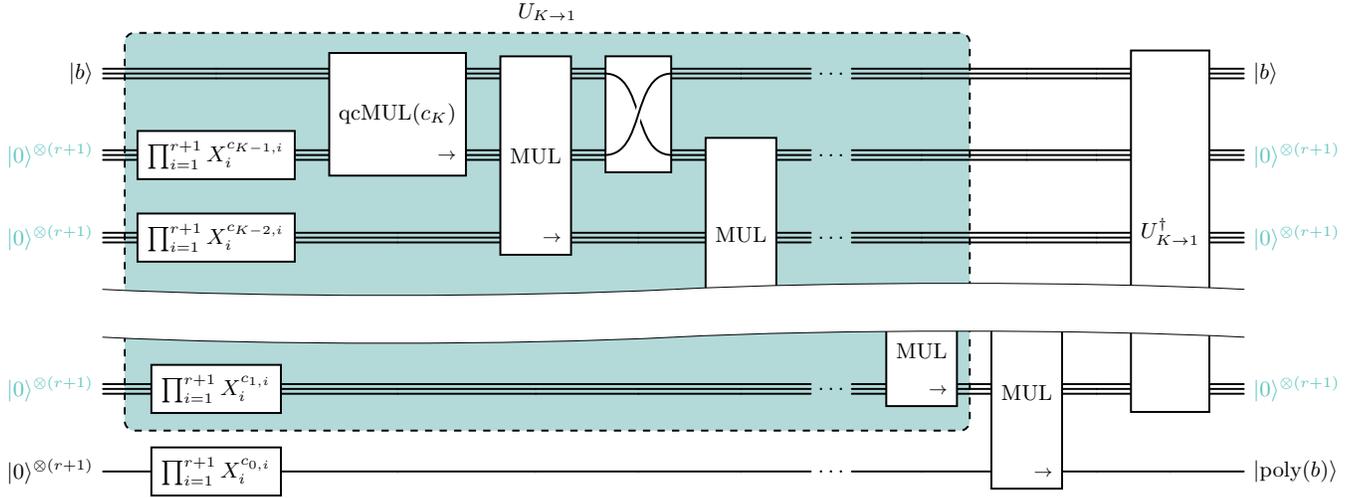

The desired quantum operation POLY shall read a number $b$ stored in a register and return $\mathrm{poly}(b)$ in another register
\begin{equation}
	\mathrm{POLY}\ket{b}\ket{0}
	=
	\ket{b}\ket{\mathrm{poly}(b)}.
\end{equation}
The full implementation of POLY is described in \cref{fig:.poly}.
It starts with encoding the coefficients $\{c_0,\dots,c_{K-1}\}$ in $K$ quantum registers next to the input register
\begin{equation}
	\ket{b,c_{K-1},c_{K-2},\dots,c_0}.
\end{equation}
Next, it computes the most inner part of \cref{eq:Horner} by computing the product $bc_K$ with the quantum-classical multiplier qcMUL and adds it to the second register.
Now, we have
\begin{equation}
	\ket{b,c_{K-1}+b c_K,c_{K-2},\dots,c_0}.
\end{equation}
A series of $K-1$ iterations follows in which the previous result is multiplied by $b$ before being added to the next register.
This gives
\begin{equation}
	\ket{b,c_{K-1}+b c_K,c_{K-2}+b(c_{K-1}+b c_K),\dots,\mathrm{poly}(b)}.
\end{equation}
At last, it remains to uncompute the intermediate results using the Hermitian conjugate of all previous gates not applied to the target register.
In this way, we get the final state
\begin{equation}
	\ket{b,0,0,\dots,\mathrm{poly}(b)}.
\end{equation}

This routine requires a total $n_{\mathrm{POLY}}$ qubits with $n_{\mathrm{POLY}}$ ancillas
\begin{subequations}
	\begin{align}
		n_{\mathrm{POLY}}
		 & =
		(K+1)(r+1)+n_{\mathrm{ADD,a}},
		\\
		n_{\mathrm{POLY,a}}
		 & =
		(K-1)(r+1)+n_{\mathrm{ADD,a}}.
	\end{align}
\end{subequations}
The runtime is dominated by the multipliers
\begin{equation}
	t_\mathrm{POLY}
	=
	\bigO(Kt_\mathrm{MUL}).
\end{equation}
Assuming non-optimal but memory efficient carry-ripple adder based multiplication we get $t_\mathrm{POLY}=\bigO(Kr^2)$.

In \cref{sec:exponentiation} we describe a more general routine for the computation of signomial functions that can surpass Horner's scheme if the total number of non-zero terms is small in comparison to the maximum exponent $K$.

\subsection{Square root with Newton-Raphson method}
\label{sec:newton-raphson}

The computation of the square root is closely related to division methods.
Here, as well, we can learn from classical algorithms.
Starting with the slow division, restoring and non-restoring versions of division have been developed for quantum devices~\cite{thapliyal2017,Gayathri2021}.
On the other hand, classical fast division methods exist such as the Goldschmidt division and the Newton-Raphson division.
The quantum version of Goldschmidt division is described in Ref.~\cite{Gayathri2021}.
In this section, we focus on the Newton-Raphson division that was already described in Refs. \cite{Haner2018,Gayathri2022Newton}, and propose an analog routine for the reciprocal and standard square root.
Furthermore, we propose an alternative version of the standard Newton-Raphson division, that reduces the memory requirements by reinitializing all ancilla registers in \cref{sec:reciprocal}.

The Newton-Raphson method allows us to find the root of a function $f(x)$ in an iterative process.
For this we approximate $f(x)$ with its tangent at a given estimate $x_k$ and determine the root $x_{k+1}$ of the tangent:
\begin{equation}
	x_{k+1}
	=
	x_k -\frac{f(x_k)}{f'(x_k)}.
	\label{eq:nr_iteration}
\end{equation}
This becomes the new estimate for the root.
For most functions, the new estimation is closer to the exact root.
One can repeat this process multiple times to reduce the error.
The convergence rate of this method depends on $f(x)$ and the initial estimate $x_0$.

\paragraph{Reciprocal square root}
The Newton-Raphson method can be used to compute the reciprocal square root $S^{-1/2}$ with $S\ge 0$.
This in turn can be further used for the computation of the regular square root $S^{1/2}$ simply by multiplying its result by $S$.
We need a function for which the root fulfills $x=S^{-1/2}$
\begin{equation}
	f_S(x)
	=
	\frac{1}{x^2}-S,
\end{equation}
and accordingly
\begin{equation}
	x_{k+1}
	=
	\frac{1}{2}x_k(3-Sx_k^2).
\end{equation}
Here, we can also define an error $\varepsilon_k=\abs{1-\sqrt{S}x_k}$ that changes like
\begin{multline}
	\varepsilon_{k+1}
	=
	\abs{1-\sqrt{S}x_{k+1}}
	=
	\\
	\abs{
		1+\frac{1}{2}\sqrt{S}x_k
	}
	\left(
	1-\sqrt{S}x_k
	\right)^2
	=
	\abs{
		1+\frac{1}{2}\sqrt{S}x_k
	}
	\varepsilon_k^2.
\end{multline}
This converges with quadratic order, if $\varepsilon_{k+1}/\varepsilon_{k}<1$, and $\varepsilon_k<1$ which is equivalent to
\begin{equation}
	\sqrt{S}x_k \in \left(\frac{-\sqrt{17} - 1}{2},-1\right)
	\cup
	\left(0,\frac{\sqrt{17} -1}{2}\right),
\end{equation}
and
\begin{equation}
	\sqrt{S}x_k \in (0,2),
\end{equation}
respectively.
Hence, the method converges as long as $\sqrt{S}x_0\in(0,1.56)$.
Due to the quadratic convergence, the number of steps $L$ scales logarithmic with the desired number of correct digits
\begin{equation}
	L=\bigO(\log_2 r).
	\label{eq:L}
\end{equation}
The exact runtime relies on a good first estimation $x_0$.
A safe way to guarantee convergence is to always choose $x_0$ close to $0$.
Alternatively, one could use a lookup table that selects the initial point depending on the range of the number $S$~\cite{flynn1970,Gayathri2022Newton}, at the price of adding controlled operations.

\begin{figure*}
	\centering
	\begin{adjustbox}{max width=\linewidth}
		\begin{quantikz}[transparent,wire types={b,b,b,b,b,b,b},classical gap=0.07cm]
			\lstick{$\ket{S}$} &\gate[3,label style={yshift=5mm}]{\mathrm{MUL}} &&&&&&&&\gate[3,label style={yshift=5mm}]{\mathrm{MUL}^\dagger}&\rstick{$\ket{S}$}
			\\
			\lstick{\blue{$\ket{0}^{\otimes r}$}} &\gateoutput{$\rightarrow$} &\gate[3]{\mathrm{MUL}}&&&&&&\gate[3]{\mathrm{MUL}^\dagger} &\gateoutput{$\rightarrow$}&\rstick{\blue{$\ket{0}^{\otimes r}$}}
			\\
			\lstick{$\ket{x_k}$} &&&&\gate[2,swap]{}&&\gate[2,swap]{}&&&&\rstick{$\ket{x_k}$}
			\\
			\lstick{$\blue{\ket{0}^{\otimes r}}$} &&\gateoutput{$\rightarrow$}&\gate[2]{\mathrm{SUB}^{2^{-1}}}&&\gate[3,label style={yshift=0mm}]{\mathrm{MUL}} &&\gate[2]{\mathrm{ADD}^{2^{-1}}}&\gateoutput{$\rightarrow$}&&\rstick{$\blue{\ket{0}^{\otimes r}}$}
			\\
			\lstick{$\blue{\ket{0}^{\otimes r}}$}&&\gate{X_{p}X_{1+p}}&\gateoutput{$\rightarrow$}&&&&\gateoutput{$\rightarrow$}&\gate{X_{p}X_{1+p}}&&\rstick{$\blue{\ket{0}^{\otimes r}}$}
			\\
			\lstick{$\ket{0}^{\otimes (r)}$} &&&&&\gateoutput{$\rightarrow$}&&&\rstick{$\ket{x_{k+1}=\frac{1}{2}x_{k}(3-Sx_k^2)}$}
		\end{quantikz}
	\end{adjustbox}
	\caption[Circuit: Newton-Raphson iteration step for the reciprocal square root]{
	Circuit for one iteration step of the quantum Newton-Raphson method tailored to find $S^{-1/2}$.
	Ancilla qubits required by MUL, $\mathrm{ADD}^{2^{-1}}$, and $\mathrm{SUB}^{2^{-1}}$ are hidden.
	Here, $\ket{S}$ could be a superposition of states allowing parallel computing.
	}
	\label{fig:nr_sqrt}
\end{figure*}
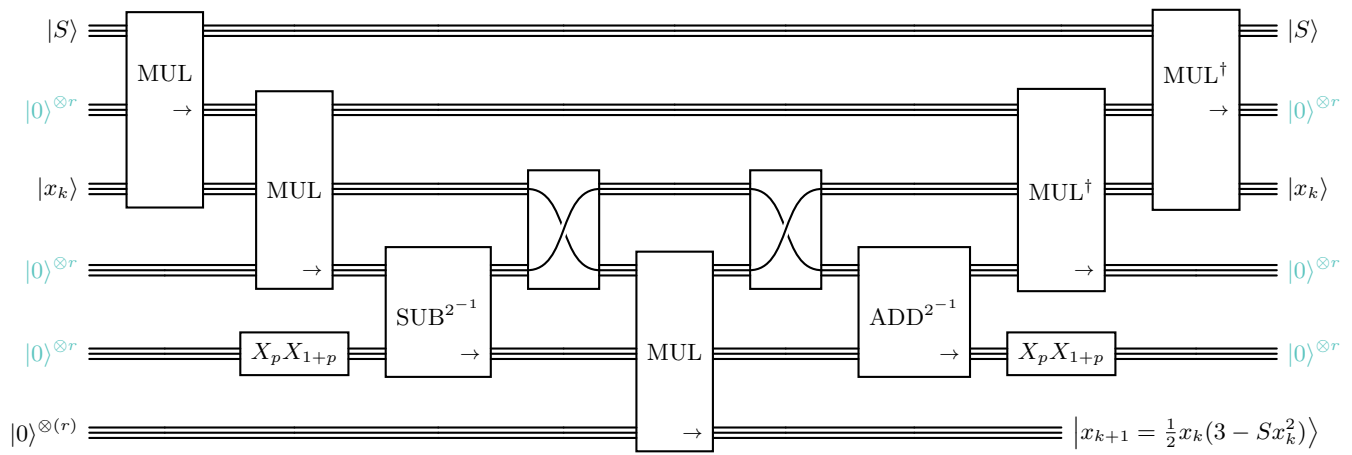

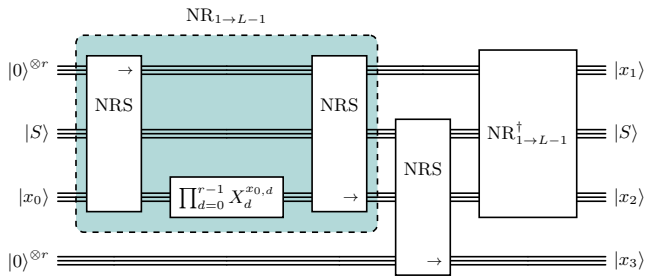
\begin{figure}
	\centering
	\begin{adjustbox}{max width=\linewidth}
		\begin{quantikz}[wire types={b,b,b,b},classical gap=0.07cm]
			\lstick{$\ket{0}^{\otimes r}$} &\gate[3,label style={yshift=5mm}]{\mathrm{NRS}}\gategroup[
				3,
				steps=3,
				style={
						dashed,
						rounded corners,
						fill=teal!30,
						inner xsep=2pt
					},
				background,
				label style={
						label position=above,
						anchor=south,
						yshift=-0.15cm
					}
			]{$\mathrm{NR}_{1\to L-1}$}\gateoutput{$\rightarrow$}&&\gate[3,label style={yshift=5mm}]{\mathrm{NRS}}&&\gate[3]{\mathrm{NR}_{1\to L-1}^\dagger}&\rstick{$\ket{x_1}$}
			\\
			\lstick{$\ket{S}$} &&&&\gate[3,label style={yshift=5mm}]{\mathrm{NRS}}&&\rstick{$\ket{S}$}
			\\
			\lstick{$\ket{x_0}$} &&\gate{\prod_{d=0}^{r-1}X_d^{x_{0,d}}}&\gateoutput{$\rightarrow$}&&&\rstick{$\ket{x_2}$}
			\\
			\lstick{$\ket{0}^{\otimes r}$} &&&&\gateoutput{$\rightarrow$}&&\rstick{$\ket{x_3}$}
		\end{quantikz}
	\end{adjustbox}
	\caption[Circuit: Newton-Raphson method]{
		Full quantum Newton-Raphson method.
		The Newton-Raphson steps (NRS) are described in \cref{fig:nr_rec}.
	}
	\label{fig:nr}
\end{figure}

A quantum implementation of the Newton-Raphson method for the implementation of the reciprocal square root was previously briefly discussed in Ref.~\cite{Haner2018}.
We show a detailed implementation of one Newton-Raphson iteration in \cref{fig:nr_sqrt}.
It starts by computing $Sx_k^2$ in two iterations of MUL.
After this, we subtract it from $1.5$, that we encoded, using two NOT gates, in an ancilla register.
Here, we use a subtractor based on the modified adder illustrated in \cref{fig:modadd_k_b}.
This allows us to subtract only half of $Sx_k^2$.
A final multiplication by $x_k$ yields in $x_{k+1} = x_k(1.5-0.5Sx_k^2)$.
These steps are repeated $L$ times as illustrated in \cref{fig:nr} for the case $L=3$.
The initial estimation $x_0$ is known, which allows us to uncompute and reuse its register.
This reduces the total memory requirements by one register.
The remaining registers with the intermediate results $\{x_1,\dots,x_{L-1}\}$ have to keep their information until the very end before we uncompute them all together.
For the square root one multiplies $x_L$ by $S$ and stores the result in an additional register.

The memory requirements of one step are
\begin{subequations}
	\begin{align}
		n_\mathrm{SQRT-step}
		 & =
		6r+n_\mathrm{ADD,a},
		\\
		n_\mathrm{SQRT-step,a}
		 & =
		3r+n_\mathrm{ADD,a}.
	\end{align}
\end{subequations}
Here, $n_\mathrm{SQRT-step}$ is the total qubit number and $n_\mathrm{SQRT-step,a}$ only the ancillas.
We assumed $r$-qubit register without sign qubits to prevent complex results ($S<0\to \sqrt{S}\neq \mathbb{R}$).
The whole Newton-Raphson method for the square root, with $L$ iterations, requires
\begin{subequations}
	\begin{align}
		n_\mathrm{SQRT}
		 & =
		(L+4)r+n_\mathrm{ADD,a}+r,
		\\
		n_\mathrm{SQRT,a}
		 & =
		(L+2)r+n_\mathrm{ADD,a}+r.
	\end{align}
\end{subequations}
Qubits in total and in ancillas respectively.
The first and second Newton-Raphson step (NRS) requires $n_\mathrm{SQRT-step}$ qubits, and after this we need to increase the qubit number by $r$ for the target of each additional Newton-Raphson step.
The last $r$ qubits are only required for the product of $S^{-1/2}$ and $S$.
The runtime of the full computation is
\begin{multline}
	t_\mathrm{SQRT}
	=
	(2L-1)\left(
	5t_\mathrm{MUL}
	+2t_\mathrm{ADD}
	\right)
	\\
	+t_\mathrm{MUL}
	=
	\bigO(
	L t_\mathrm{MUL}
	).
\end{multline}
Here, we omit the runtimes of the encoding of $x_0$ and the SWAP gates due to their small impact next to $t_\mathrm{MUL}$.
The additional $t_\mathrm{MUL}$ is required for the final product $S^{-1/2}S$, but is not changing the overall scaling.
We do not replace $L$ using \cref{eq:L}, because its optimum depends on the difference between the desired and initial number of correct digits and with it usually performs better than $\log_2 r$.

\section{Stiffness and mass matrix for continuous problems}
\label{sec:cont_prob}

The derivation of the mass and stiffness matrices for a continuous, solid structure using \ac{fem} is well understood and documented in the literature (e.g. \cite{Ern2004,rao2005,fish2007,cook1989}).
The derivation of force equations in continuous systems can be intricate and less intuitive, which is why we describe an alternative more concise derivation in \cref{sec:osci} based on the Lagrangian formalism.
We further give an introduction to test functions and to \ac{fem} applied to classical oscillator problem.
\cref{sec:limits_tf} describes the limitations for the choice of test function.
\cref{sec:mass_lumping,sec:stiffness_1d} contain simplifications of the mass and stiffness matrix, respectively, that allow us to compute them efficiently on a quantum computer as described in \cref{sec:application}.

\subsection{Elastic structures}
\label{sec:osci}
We consider a continuous solid in $d$ dimensions defined in a domain $\Omega \subseteq \mathbb{R}^d$, whose dynamical behavior is described by its kinetic and potential energy density.
We combine both of them in the Lagrangian density
\begin{multline}
	l(\bm{x}, t)
	=
	\frac{1}{2}\rho(\bm{x}) \bm{v}(\bm{x}, t)^T \bm{v}(\bm{x}, t)
	\\
	-
	\frac{1}{2}\sum_{\alpha, \beta,\gamma,\delta=1}^d
	Y_{\alpha\beta\gamma\delta}(\bm{x})\varepsilon_{\alpha\beta}(\bm{x}, t)\varepsilon_{\gamma\delta}(\bm{x}, t).
	\label{eq:strong_form}
\end{multline}
The first term is the kinetic energy per unit of volume with the density $\rho$ and the velocity $\bm{v}=(v_{1},\dots,v_{d})^T\in\mathbb{R}^{d}$ (cf. \cite{bampi1983, cook1989}).
The potential is reduced to its contribution from the elastic energy, neglecting others like electromagnetic or gravitational potentials.
Hence, the second term in \cref{eq:strong_form} is the elastic energy per unit volume, with the material-dependent elasticity tensor $Y$ and the strain tensor $\varepsilon$.
We recommend Refs.~\cite{Ern2004, fish2007, holzapfel2002, Nye1985, Dove1993, Han2009, huang2018, sadd2014, bower2010} for a detailed derivation and discussion of the symmetry properties of $Y$ and $\varepsilon$.
For now, we are only interested in the following property
\begin{equation}
	Y_{\alpha\beta\gamma\delta}
	=
	Y_{\alpha\beta\delta\gamma}
	=
	Y_{\beta\alpha\delta\gamma}.
	\label{eq:Y_sym}
\end{equation}
In its general form, all contributions in \cref{eq:strong_form} depend on the position $\bm{x}=(x_{1},\dots,x_{d})^T\in\mathbb{R}^d$ in space and on the time $t$.
For the sake of compactness, we mostly omit the dependency on position $\bm{x}$ and time $t$, unless needed for clarity.
Both the strain tensor and the velocity field are defined in terms of the material displacement $\bm{u}=(u_{1},\dots,u_{d})^T$ from its equilibrium position
\begin{subequations}
	\begin{align}
		\varepsilon_{\alpha\beta}
		 & =
		\frac{1}{2}\left(
		\parderiv{u_{\alpha}}{x_{\beta}}
		+\parderiv{u_{\beta}}{x_{\alpha}}
		\right),
		\label{eq:strain}
		\\
		\bm{v}
		 & =
		\frac{\partial \bm{u}}{\partial t}.
		\label{eq:velocity}
	\end{align}
\end{subequations}
The strain describes the change of the displacement with respect to the position $\bm{x}$ and the velocity field the analogue with respect to time $t$.
The total Lagrangian of the system reads
\begin{multline}
	\mathcal{L}
	=
	\frac{1}{2}\int_{\mathbb{R}^d} \Diff{d} x \, \rho \bm{v}^T \bm{v}
	\\
	-
	\frac{1}{2}\sum_{\alpha, \beta,\gamma,\delta=1}^d\int_{\mathbb{R}^d} \Diff{d} x \,
	Y_{\alpha\beta\gamma\delta}
	\parderiv{u_{\alpha}}{x_{\beta}}
	\parderiv{u_{\gamma}}{x_{\delta}},
	\label{eq:tot_lagr}
\end{multline}
where we used \cref{eq:strain,eq:Y_sym} to replace $\varepsilon_{\alpha\beta}$.
The corresponding Euler-Lagrange equation describes the motion of the system.
In finite element analysis, the continuous equation of motion is called the \emph{strong form} because it is valid at every point $\bm{x}$.
Solving a strong form analytically is not always possible which is why numerical approaches are common in classical simulations.
For this, we discretize \cref{eq:tot_lagr} and with it its Euler-Lagrange equation, obtaining a system of equations that reproduces the exact solution at a finite number of points $N$.
On one hand, the number $N$ should be small so that a computer can solve the equation for all those points in reasonable time.
On the other hand, it should be large enough to approximate the behavior of the full system sufficiently.
Finding the sweets-spot can be difficult in practice and gives rise to a trade-of between accuracy and computation cost.

In order to discretize the system, we replace $\bm{u}(\bm{x})$ with an approximation $\bar{\bm{u}}(\bm{x})$
\begin{equation}
	\bar{\bm{u}}(\bm{x})
	=
	\sum_{i=1}^N \bar{\bm{u}}_i \varphi_i(\bm{x}),
	\qquad \bar{\bm{u}}_i = \bm{u}(\bm{x}_i).
	\label{eq:u_app}
\end{equation}
$\bar{\bm{u}}(\bm{x})$ is a linear combination of so-called \emph{test functions} $\varphi_i(\bm{x})$ which are only non-zero within small parts of the domain.
This approximation is exact at $N$ arbitrary positions $\bm{x}\in\{\bm{x}_1,\dots,\bm{x}_N\}$\footnote{
	The Roman subscripts here count over the $N$ positions and should not be confused with the Greek superscripts that count over the
	$d$ spatial dimensions.
}.
These positions should be well distributed within the geometry $\Omega=\{\bm{x}:\rho(\bm{x}),Y_{\alpha\beta\gamma\delta}(\bm{x})\neq 0\}$ in order to obtain a good approximation.
However, they are not limited to $\Omega$.
For the purpose of a simpler mesh one can place some points outside $\Omega$.
A similar approximation $\bar{\bm{v}}(\bm{x})$ is required for the velocity field
\begin{equation}
	\bar{\bm{v}}(\bm{x})
	=
	\sum_{i=1}^N \bar{\bm{v}}_i \phi_i(\bm{x}),
	\qquad
	\bar{\bm{v}}_i = \bm{v}(\bm{x}_i) = \frac{\mathrm{d} \bm{u}_i}{\mathrm{d} t}.
	\label{eq:v_app}
\end{equation}
Here, $\{\phi_i(\bm{x}),\forall i \in [N]\}$ is another set of test functions, which are not necessarily equal to $\varphi_i(\bm{x})$.
Using the two approximations \cref{eq:u_app,eq:v_app} in \cref{eq:tot_lagr}
yields the discretized Lagrangian
\begin{equation}
	\bar{\mathcal{L}} 
	=
	\frac{1}{2}  \sum_{i,j=1}^{N} \sum_{\alpha, \gamma=1}^d 
	\left[ v_{i \alpha}  M_{i\alpha, j \gamma} v_{j\gamma}
	-
	u_{i\alpha}
	K_{i \alpha, j \gamma}
	u_{j \gamma} \right].
	\label{eq:lagrangian}
\end{equation}
In \cref{eq:lagrangian} we introduced the elements of the $Nd \times Nd$ mass matrix $\bm{M}$ defined as
\begin{equation}
	M_{i\alpha, j \gamma}
	=
	\delta_{\alpha\gamma}\int_{\mathbb{R}^d} \Diff{d}x \, \rho \phi_i \phi_j .
	\label{eq:mass_matrix}
\end{equation}
with $\delta_{\alpha\gamma}$ the Kronecker delta.
Similarly, the elements of the $Nd \times Nd$ stiffness matrix $\bm{K}$ are defined as
\begin{equation}
	K_{i\alpha, j \gamma}
	=
	\sum_{\beta,\delta=1}^d
	\int_{\mathbb{R}^d} \Diff{d}x \,
	Y_{\alpha\beta\gamma\delta}
	\parderiv{\varphi_i}{x_{\beta}}
	\parderiv{\varphi_j}{x_{\delta}}
	.
	\label{eq:stiffness_matrix}
\end{equation}
Denoting by $\hat{e}_{i \alpha}$ one of $Nd$-dimensional unit, column vectors, the mass and stiffness matrices can formally be written as
\begin{subequations}
	\label{eq:massstiffmat}
	\begin{align}
		\bm{M}
		 & =
		\sum_{i,j=1}^{N}\sum_{\alpha,\gamma=1}^d
		M_{i\alpha, j \gamma}
		\hat{e}_{i \alpha} \hat{e}_{j \gamma}^T,
		\\
		\bm{K}
		 & =
		\sum_{i,j=1}^{N}\sum_{\alpha,\gamma=1}^d
		K_{i\alpha, j \gamma}
		\hat{e}_{i \alpha}\hat{e}_{j \gamma}^T,
	\end{align}
\end{subequations}
respectively.
Note that the mass and stiffness matrix are both symmetric, i.e., $M_{i \alpha, j \gamma} = M_{j\gamma, i \alpha}$ and $K_{i\alpha, j\gamma} = K_{j \gamma, i \alpha}$.
We now collect all dynamical variables $\bar{u}_{i \alpha}$ in a single $Nd$-dimensional column vector defined as
\begin{equation}
	\vectorarrow{u}
	=
	\sum_{i=1}^{N}\sum_{\alpha=1}^{d}
	\bar{u}_{i \alpha} \hat{e}_{i \alpha}.
	\label{eq:dN_u}
\end{equation}
With these definitions, the Lagrangian $\bar{\mathcal{L}}$ in \cref{eq:lagrangian} can be written compactly as
\begin{equation}
	\bar{\mathcal{L}}
	=
	\frac{1}{2} \frac{\mathrm{d} \vectorarrow{u}}{\mathrm{d} t}^T M \frac{\mathrm{d} \vectorarrow{u}}{\mathrm{d} t}
	- \frac{1}{2} \vectorarrow{u}^T K \vectorarrow{u}.
\end{equation}
Using the symmetry properties of $\bm{M}$ and $\bm{K}$, the associated Euler-Lagrange equations can be compactly written as
\begin{equation}
	\bm{M} \frac{\mathrm{d}^2 \vectorarrow{u}}{\mathrm{d} t}
	=
	- \bm{K} \vectorarrow{u}.
	\label{eq:eom}
\end{equation}
This is equivalent to \cref{eq:generalized_epp} after introducing $\vectorarrow{z}=\bm{M}^{1/2}\vectorarrow{u}$ and applying $\bm{M}^{-1/2}$ from the left.

\subsection{Limitations of the test functions}
\label{sec:limits_tf}

In general an arbitrary non-zero test function can be chosen as long as \cref{eq:mass_matrix,eq:stiffness_matrix} are well-defined.
Furthermore, the $i$-th test function has to satisfy
\begin{equation}
	\varphi_i(\bm{x}_j)
	=
	\delta_{ij},
	\label{eq:lim_exact_solutions}
\end{equation}
with $\delta_{ij}$ the Kronecker delta, to guarantee $\bar{\bm{u}}(\bm{x}_i)= \bm{u}(\bm{x}_i) = \bar{\bm{u}}_i, \, \forall i$.

The approximation $\bar{\bm{u}}(\bm{x})$ does not need to be exact for positions $\bm{x}\notin\{\bm{x}_1.\dots,\bm{x}_N\}$.
However, they are conventionally chosen close to the exact solution $\bm{u}(\bm{x})$.
Hence, $\varphi_i(\bm{x})$ should fulfill
\begin{equation}
	\sum_{i}^{N}\varphi_i(\bm{x})
	\approx 1,
	\qquad
	\forall \bm{x}.
\end{equation}
The same considerations hold for $\phi_i(\bm{x})$ as well.

The number of overlapping test functions at a point $\bm{x}$ defines the sparsity of our matrices $\bm{K}$ and $\bm{M}$.
Sparse matrices are easier to handle in numerical calculations, which is why another convention is to set the test function to zero for large parts of the domain.

\subsection{Boundary conditions}
\label{sec:bc}
Up to this point we have considered a fully free system, i.e. without any boundary conditions.
For structural problems it is common to fix a given number of nodes $\bm{x}_{i}\in\mathcal{X}_\mathrm{D}$ (i.e. $u_{i\alpha}=\mathrm{const}$), which is called Dirichlet boundary condition.
Due to this, we know some \emph{unknown} in $\vectorarrow{u}$ in advance which allows us to rearrange the original equation of motion (see \cref{eq:eom,eq:generalized_epp}), which we extend by a force term
\begin{equation}
	\left(
		\bm{H}
		+\frac{\Diff{2}}{\diff t^2}
	\right)\vectorarrow{z}
	=
	\vectorarrow{f},
\end{equation}
with $\vectorarrow{z}=\bm{M}^{1/2}\vectorarrow{u}$.
We sort the elements of $\vectorarrow{z}$ into \emph{unknown} $\vectorarrow{z}_\mathrm{u}$ and \emph{known} $\vectorarrow{z}_\mathrm{k}$.
This allows us to write
\begin{equation}
	\begin{pmatrix}
		\bm{H}_\mathrm{kk}+\frac{\Diff{2}}{\diff t^2} & \bm{H}_\mathrm{ku}
		\\
		\bm{H}_\mathrm{uk} & \bm{H}_\mathrm{uu}+\frac{\Diff{2}}{\diff t^2}
	\end{pmatrix}
	\begin{pmatrix}
		\vectorarrow{z}_\mathrm{k}
		\\
		\vectorarrow{z}_\mathrm{u}
	\end{pmatrix}
	=
	\begin{pmatrix}
		\vectorarrow{f}_\mathrm{k}
		\\
		\vectorarrow{f}_\mathrm{u}
	\end{pmatrix}.
\end{equation}
Applying force to a fixed node is useless (i.e. $\vectorarrow{f}_\mathrm{k}=0$), thus the latter reduces to our new equation of motion
\begin{equation}
	\left(
		\bm{H}_\mathrm{uu}
		+\frac{\Diff{2}}{\diff t^2}
	\right)
	\vectorarrow{z}_\mathrm{u}
	=
	\vectorarrow{f}_\mathrm{u}-\bm{H}_\mathrm{uk}\vectorarrow{z}_\mathrm{k}.
	\label{eq:red_eom}
\end{equation}
Hence, applying Dirichlet boundary conditions is equivalent to reducing the matrix $\bm{H}$ to the remaining free nodes followed by a redefinition of the force term $\vectorarrow{f}_\mathrm{u}\to\vectorarrow{f}_\mathrm{u}-\bm{H}_\mathrm{uk}\vectorarrow{z}_\mathrm{k}$.
Other boundary conditions (e.g. Neumann boundary conditions) effect also only the force term.
We recommend Refs.~\cite{rao2005,fish2007} for more details
For the computation of the eigenpairs we focus on the harmonic part of \cref{eq:red_eom} in the following.

\subsection{Mass lumping}
\label{sec:mass_lumping}
\begin{figure}
	\subfloat[Box test function]{\label{fig:1d_tf_mass}
		\centering
		\includegraphics[width=6.5cm]{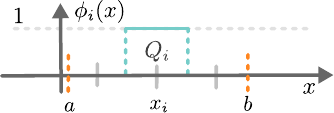}
	}

	\subfloat[Triangular test function]{\label{fig:1d_tf}
		\centering
		\includegraphics[width=6.5cm]{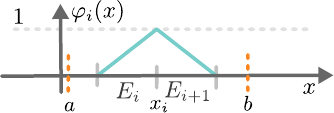}
	}
	\caption{
		(a) One-dimensional test function $\phi_i(x)$ (in \blue{blue}) for the mass matrix in an equidistant mesh.
		In \orange{orange}, we see the limits $a$ and $b$ of the geometry $\Omega$.
		(b) One-dimensional test function $\varphi_i(x)$ (\blue{blue}) for the stiffness matrix in an equidistant mesh.
		In \orange{orange}, we see the limits $a$ and $b$ of the geometry $\Omega$.
	}
\end{figure}

\begin{figure}
	\centering
	\includegraphics[width = 6cm]{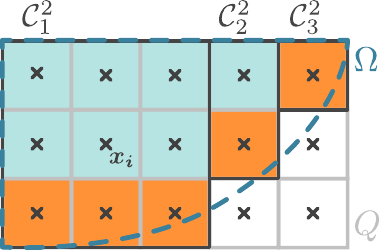}
	\caption{
		Alignment between the geometry $\Omega$ within the \darkblue{dark blue} dashed line and the elements $Q_i\subset Q$ shown as a grid.
		The elements in \blue{light blue} are fully part of $\Omega$ and with it fulfil $\rho\neq 0$.
		In \orange{orange} ($\bm{x}_{\bm{i}}\in\Omega$) we extend $\rho$ to the whole element to preserve the symmetry.
		White elements have $\rho=0$.
		A simplified geometry $\Omega'=\bigcup_{i=1}^3\mathcal{C}^2_i$ is shown in black contours
	}
	\label{fig:geometry}
\end{figure}

From \cref{eq:mass_matrix}, it is clear that one can always choose the test functions $\phi_i(\bm{x})$ so that the mass matrix is positive and diagonal (a beneficial form, cf. \cref{sec:application}).
One way of achieving this is by ``mass lumping'', which is a well established technique \cite{felippa2004, cook1989}.
In what follows, we
describe one version that relies on box-functions for $\phi_i(\bm{x})$.
Mathematically, this is achieved by first introducing a set $Q \subset \mathbb{R}^d$ such that $\Omega \subseteq Q$.
We partition $Q$ into $N$ subdomains $Q_i$ such that $\bm{x}_i\in Q_i$, $Q = \cup_{i=1}^N Q_i$, and $Q_i \cap Q_j = \emptyset,\forall i \neq j$.
Examples for one-dimensional and two-dimensional geometries are shown in \cref{fig:1d_tf_mass} and \cref{fig:geometry} respectively.
We can then take
\begin{equation}
	\phi_i(\bm{x})
	=
	\begin{cases}
		1 & \quad \mathrm{ if~} \bm{x}\in Q_i,
		\\
		0 & \quad \mathrm{ if~} \bm{x}\notin Q_i.
	\end{cases}
\end{equation}
With this choice the elements of the mass matrix become
\begin{equation}
	M_{i\alpha, j \gamma}
	=
	\delta_{\alpha\gamma}\delta_{ij}\int_{Q_i}
	\Diff{d}x \, \rho
\end{equation}

In order to gain further insights into the calculations that need to be performed, let us calculate explicitly the matrix elements $M_{i\alpha, j \gamma}$ for a $d$-dimensional hypercuboid mesh $\mathcal{C}^d$ with constant grid size $\Delta$ (see~\cref{fig:geometry} for an example in two dimensions).
For homogeneous systems with constant density $\rho$, and extending $\rho$ to all $Q_i$ for which $\bm{x}_i\in\Omega$ (keeping $\rho=0$ in $Q_i$ for which $\bm{x}_i\notin\Omega$), the elements of the mass matrix are
\begin{equation}
	M_{i\alpha, j \gamma}
	=
	\begin{cases}
		\delta_{\alpha\gamma}\delta_{ij} \rho \Delta^d &
		\mathrm{if~} \bm{x}_i \in \Omega,
		\\
		0                                                & \mathrm{otherwise}.
	\end{cases}
	\label{eq:mass_homo}
\end{equation}

The integration is more involved for non-constant densities, which are either polynomial or can be approximated by polynomials in $\bm{x}$.
Typically, for regular meshes, they lead to matrix elements that are polynomials in the label $i$.

\subsection{Stiffness matrix entries in one dimension}
\label{sec:stiffness_1d}

In this section, we simplify the elements of the stiffness matrix $\bm{K}$ (see \cref{eq:stiffness_matrix}).
For illustrative reasons, we limit our analysis to the one-dimensional case, in which the domain can simply be taken as an interval $\Omega = [a, b]$ with $a<b$.
Non-connected geometries can be considered as separate problems.
Setting $Y_{1111}\to Y$ and assuming a uniform structure with constant $Y$ within $\Omega$, we have
\begin{equation}
	K_{ij}^{(\mathrm{1D})}
	=
	Y \int_\Omega \diff x \,	
	\parderiv{\varphi_i}{x}
	\parderiv{\varphi_j}{x}
	.
	\label{eq:1d_stiffness_matrix}
\end{equation}
To discretize the problem we introduce $N$ points $\{x_1, x_2, \dots, x_N\}$ such that $[x_1, x_N]\subseteq [a, b]$ with $x_1 < x_2 < \dots < x_N$.
These points align with those used in the mass matrix.
In one dimension, we can always choose a set of points that is uniformly distributed with gap size $\Delta = x_{i+1}-x_i ,\forall i\in [N-1]$.
Here we chose one for which the lower (higher) border of the one dimensional elements $Q_1^\mathrm{(1D)}$ ($Q_N^\mathrm{(1D)}$) aligns with $a$ ($b$), i.e. $x_i=a+(i-1/2)\Delta$ with $\Delta=(b-a)/N$ (see \cref{fig:1d_tf_mass,fig:1d_tf}).
We introduce $N$ linear test functions defined as
\begin{equation}
	\varphi_i(x)
	=
	\begin{cases}
		1+\frac{x-x_i}{\Delta} & \mathrm{if~}  x \in E_i,
		\\
		1-\frac{x-x_i}{\Delta} & \mathrm{if~}  x \in E_{i+1},
		\\
		0                      & \mathrm{otherwise},
	\end{cases}
	\label{eq:tf_equidist_mesh}
\end{equation}
with $i \in [N]$ and $E_i=\left(x_i- \Delta,x_{i}\right]$.
$E_i$ and $E_{i+1}$ are the two sub-intervals on the left and on the right of node $x_i$.
Note that the test functions are overlapping as they are non-zero in two sub-intervals.
\cref{eq:tf_equidist_mesh} is illustrated in \cref{fig:1d_tf}.

The stiffness matrix (\cref{eq:1d_stiffness_matrix}) requires the first derivatives of the test function
\begin{equation}
	\parderiv{\varphi_i}{x}
	=
	\begin{cases}
		\frac{1}{\Delta}  & \mathrm{if~}  x\in E_i,
		\\
		-\frac{1}{\Delta} & \mathrm{if~}  x\in E_{i+1},
		\\
		0                 & \mathrm{otherwise},
	\end{cases}
	\label{eq:derivative_tf}
\end{equation}
and finally the product of two of them
\begin{equation}
	\parderiv{\varphi_i}{x}
	\parderiv{\varphi_j}{x}
	=
	\begin{cases}
		\delta_{ij}\frac{1}{\Delta^2}
		-\delta_{i-1,j}\frac{1}{\Delta^2} &
		\mathrm{if~} (x \in E_i),
		\\
		\delta_{ij}\frac{1}{\Delta^2}
		-\delta_{i+1,j}\frac{1}{\Delta^2} &
		\mathrm{if~} (x\in E_{i+1}),
		\\
		0                                 & \mathrm{otherwise}.
	\end{cases}
	\label{eq:tf_product}
\end{equation}
The matrix elements of the stiffness matrix read
\begin{multline}
	K_{ij}^{(\mathrm{1D})}
	=
	\left(
		\delta_{ij}
		-\delta_{i-1,j}
	\right)\frac{Y}{\Delta^2}\int_{E_i\cap\Omega}\diff x
	\\
	+\left(
		\delta_{ij}
		-\delta_{i+1,j}
	\right)\frac{Y}{\Delta^2}\int_{E_{i+1}\cap\Omega}\diff x.
	\label{eq:1d_stiffness_matrix2}
\end{multline}
The symmetry is broken only by the contour of $\Omega$.
Similar to \cref{sec:mass_lumping} we reduce the domain $\Omega$, in which $Y$ is non-zero, to every $E_i\subset \Omega$, i.e. set $E_1=E_{N+1}=\emptyset$.
The stiffness matrix for this special case splits into two parts
\begin{equation}
	K_{ij}^{(\mathrm{1D})}
	=
	\begin{cases}
		2\delta_{ij}\frac{Y}{\Delta}
		-\delta_{i\pm 1,j}\frac{Y}{\Delta} &
		\mathrm{if~} i \in [N]\setminus \{1,N\},
		\\
		\delta_{ij}\frac{Y}{\Delta}
		-\delta_{i\pm 1,j}\frac{Y}{\Delta} &
		\mathrm{if~} i\in \{1,N\},
	\end{cases}
	\label{eq:1d_stiffness_matrix3}
\end{equation}
where $[N]\setminus \{1,N\}$ is the bulk of the geometry and $\{1,N\}$ denotes the edge or boundary.

The integral can still be simplified for non-homogenous systems ($Y(x)\neq \mathrm{const.}$) if the elasticity is a polynomial in $x$.
However, this replaces the piecewise values in $K_{ij}^\mathrm{(1D)}$ with polynomials in $i$ and $j$.
Non-polynomial elasticities can be approximated by polynomials.
For problems, living in $d$ dimensions, the list of conditions in \cref{eq:1d_stiffness_matrix3} extends with the number of element types (bulk, face, edge or corner elements).
Also, the number of Kronecker-delta terms increases with the number of ways two test functions can overlap.
Furthermore, in $d$ dimensions it is not generally given that one can choose a mesh for which $Q=\Omega$.
For symmetry reasons the mesh and with it the elements will always fill a hypercuboidal space, while the given geometries $\Omega$ are usually more complex in their form.
Hence, on has to extend the list of conditions by $\bm{x}_{\bm{i}},\bm{x}_{\bm{j}}\in\Omega$ to test if a given element is contained in $\Omega$ and return $0$ otherwise (cf. \cref{eq:mass_homo}).

The computation of the stiffness matrix requires logic operations to distinguish between the different cases.
This is described in \cref{sec:logic_operator}.
In the case of non-homogeneous elasticity, it needs a routine for the calculation of polynomials.
The memory and runtime optimized computation of polynomials on quantum hardware is described in \cref{sec:compl}.

\section{Application in the calculation of response functions}
\label{sec:application}

Oracles, requiring extensive arithmetic, appear in many quantum algorithms.
Here, we discuss the implementation of the oracle appearing in the algorithm for the computation of response functions of coupled oscillators~\cite{danz2024calculating} defined in~\cref{eq:oracle}, using the quantum arithmetic discussed in~\cref{sec:compl}.
The core of this algorithm is the use of quantum phase estimation as an eigenpair solver for the matrix $\bm{H}$ (see \cref{eq:generalized_epp}).
However, due to the possible zero-rows in the mass and stiffness matrices the equation of motion \cref{eq:eom} contains trivial rows  not contributing to the overall solutions, but making $\bm{M}$ non-invertible.
We prevent this by marking those rows with flags ($\mathcal{F}_M,\mathcal{F}_K \neq 0$) on the diagonal elements.
In this way, they stay disconnected from the remaining matrix and with it contain their eigenvalues, besides adding the degenerate value $\mathcal{F}_M$ or $\mathcal{F}_K$, respectively.
For the combined matrix $\bm{H}$ (see \cref{eq:generalized_epp}) with the elements $H_{uv}=H_{i \alpha, j \gamma}$ this gives
\begin{equation}
	H_{i \alpha,j \gamma}
	=
	\begin{cases}
		\frac{K_{i\alpha, j \gamma}}{\sqrt{
				M_{i \alpha, i \alpha}
				M_{j \gamma, j \gamma}
			}}
		 & \mathrm{if~} \bm{x}_i,\bm{x}_j\in\Omega\setminus \mathcal{X}_\mathrm{D},
		\\
		\delta_{ij}\delta_{\alpha\gamma}
		\mathcal{F}
		 & \mathrm{otherwise}.
	\end{cases}
	\label{eq:H_d}
\end{equation}
Here, we used \cref{eq:mass_homo,eq:1d_stiffness_matrix3} and combined the two flags in $\mathcal{F} = \mathcal{F}_K /\mathcal{F}_M$ marking the degenerate eigenvalue, which is not contributing to the solution.
At this step, we also apply the Dirichlet boundary condition based matrix reduction (see \cref{sec:bc}) by marking the corresponding nodes $\bm{x}_i\in\mathcal{X}_\mathrm{D}$ with the flag $\mathcal{F}$.
For the use in quantum computing it is commonly required to have $N=2^n$ for some integer qubit number $n$.
The flag $\mathcal{F}$ can also be used to fill up empty rows to achieve the latter.

For a homogeneous one-dimensional system the mass and stiffness matrix are described by \cref{eq:mass_homo,eq:1d_stiffness_matrix3} thus we have
\begin{equation}
	H_{ij}^{(\mathrm{1D})}
	=
	\begin{cases}
		2\delta_{ij}\frac{Y}{\rho\Delta^2}
		-\delta_{i\pm 1,j}\frac{Y}{\rho\Delta^2} &
		\mathrm{if~} i \in [N]\setminus \{1,N\}
		\\
		&\quad\mathrm{and~} x_i,x_j\notin \mathcal{X}_\mathrm{D},
		\\
		\delta_{ij}\frac{Y}{\rho\Delta^2}
		-\delta_{i\pm 1,j}\frac{Y}{\rho\Delta^2} &
		\mathrm{if~} i\in\{1,N\}
		\\
		&\quad\mathrm{and~} x_i,x_j\notin \mathcal{X}_\mathrm{D},
		\\
		\delta_{ij}\mathcal{F}                   & \mathrm{otherwise},
	\end{cases}
	\label{eq:H_1d}
\end{equation}
with a max norm of $\norm{H}_\mathrm{max}=\max \{2Y/(\rho \Delta^{2}),\mathcal{F} \}$.
We can separate $\mathcal{F}$ from the eigenvalues by setting it larger than the largest eigenvalue  $\lambda_\mathrm{max}$
\begin{equation}
	\mathcal{F}
	>
	\norm{H}_1
	\geq
	\lambda_\mathrm{max}.
\end{equation}
In our example, the 1-norm is given by
\begin{equation}
	\norm{H}_1
	=
	\max_i\sum_{i=1}^N \abs{H_{ij}}
	=
	\frac{3Y}{\rho \Delta^{2}}.
\end{equation}

\subsection{Contour of the geometry}
\label{sec:contour}
From~\cref{eq:H_d}, we need to test if $\bm{x}_i \in \Omega\setminus\mathcal{X}_\mathrm{D}$.
In practices, the boundary conditions are usually defined as a subdomain $\mathcal{X}_\mathrm{D}\subset \Omega$ instead of a list of nodes, which is why we treat it as such in the following.
This means testing a list of conditions that need to be fulfilled before we assign any contribution to $H_{uv}$.
In this section we concentrate on the implementation of those conditions.

Many geometries can be approximated by a combination of hypercuboids (see \cref{fig:geometry}).
However, this method can be adjusted to arbitrary contours that are described by a function (see \cref{sec:adv_contours}).
Hypercuboids are defined by
\begin{equation}
	\mathcal{C}^d
	\\
	=
	\left\{
	\bm{x}:
	x_{\alpha}
	\in
	\left[
	a_{\alpha},
	b_{\alpha}
	\right],
	\forall \alpha\in\lownat{d}
	\right\},
\end{equation}
where $\alpha$ counts over all or a subset of the $d$ dimensions which are limited by the upper and lower boundaries $a_{\alpha}$ and $b_{\beta}$, respectively.
The corresponding conditions for $\bm{x}_{\bm{i}} \in \mathcal{C}^d$ are
\begin{equation}
	a_{\alpha}\leq x_{i_{\alpha},\alpha} \leq b_{\alpha},
	\quad
	\forall \alpha\in\lownat{d}.
	\label{eq:cube_conditions}
\end{equation}
Note that we remapped the index $i \to \bm{i}=(i_{1},\dots,i_{d})^T$.
This gives a total of $2d$ conditions per $\bm{x}_{\bm{i}}\in \mathcal{C}^d$.
The quantum implementation of the \emph{greater than} comparison is described in \cref{sec:compare}, while the combination of multiple conditions via \emph{logical} AND is described in \cref{sec:truth}.
For larger geometries one can extend the approximated geometry
\begin{equation}
	\Omega'
	=
	\mathcal{C}^d_1 \cup \mathcal{C}^d_2
	\cup \dots \cup
	\mathcal{C}^d_{N_\mathrm{geo}},
\end{equation}
where $\mathcal{C}^d_i$ is the $i$-th hypercuboid with unique limits and $N_\mathrm{geo}$ is the number of combined primitive geometries.
Effectively, this extends the number of conditions for $\bm{x}_{\bm{i}}\in\Omega'$ to $2dN_\mathrm{geo}$.
The conditions for $\bm{x}_{\bm{i}}\notin\mathcal{X}_\mathrm{D}$ can be constructed analogously using $N_\mathrm{D}$ hypercuboids.
In one dimension, no approximation is necessary.
$\Omega$ can be described precisely with $N_\mathrm{geo}=1$ interval $\mathcal{C}^1$.

\subsection{Required arithmetic operations}
\label{sec:complexity}

In this section, we discuss the arithmetic operations necessary for $O_\vartheta$ introduced in \cref{eq:angle,eq:oracle}.
A full list of steps is shown in \cref{alg:oracle} for a homogeneous one-dimensional problem.
\begin{algorithm}[t]
	\DontPrintSemicolon
	\caption{
		Oracle $O_\vartheta$ for $H_{ij}$ access (see~\cref{eq:oracle}) in one dimension.
		Here, we access the reduced $H'_{ij} = H_{ij}/\norm{H}_\mathrm{max}$ with flag $\mathcal{F}=4Y/(\rho \Delta^{2})$.
		CSM is the transform between the two's complement and the sign magnitude representation (see \cref{fig:circ_mult}).
		The \texttt{If} statements represent controlled operations.
	}
	\label{alg:oracle}
	\KwData{$\ket{i,j}$, $N$, and $\mathcal{X}_\mathrm{D}$}
	\KwResult{$\ket{\sgnb{H_{ij}},\vartheta_{ij}}$}
	\Begin{
	$H'\gets \ket{0}^{\otimes (r+1)}$\;
	$S \gets \ket{0}^{\otimes r}$\;
	$\vartheta \gets \ket{0}^{\otimes r}$\;
	\uIf{$x_i,x_j\in\mathcal{X}_\mathrm{D}$}{
		$H'
		\gets
		\ket{
			\delta_{i j }
		}$ \tcp*{flag}
	}
	\uElseIf{$i\in\{1,N\}$}{
		$H'
		\gets
		\ket{
			\frac{1}{4}\delta_{ij}
			-\frac{1}{4}\delta_{i\pm 1,j}
		}$ \tcp*{edge}
	}
	\Else{
		$H'
			\gets
			\ket{
				\frac{1}{2}\delta_{ij}
				-\frac{1}{4}\delta_{i\pm 1,j}
			}$ \tcp*{bulk}
	}
	$H'
		\gets
		\ket{\sgnb{H_{ij }},\abs{H'_{ij }}}
		=
		\mathrm{CSM}\ket{H'_{ij }}$\;
	$S
		\gets
		\mathrm{SQRT}\ket{\abs{H'_{ij }}}$\;
	$\vartheta
		\gets
		\mathrm{POLY}\ket{S_{ij }}$\;
	\Return $\ket{\sgnb{H_{ij }},\vartheta_{ij }}$
	}
\end{algorithm}
Extending the following for $d$ dimensions would mostly change the number of conditions that need to be tested and due to $d=\mathrm{const.}$, we neglect it from the beginning in this complexity analysis.
We further assume that $\Omega$ consists of $N_\mathrm{geo}=1$ intervals while keeping $N_\mathrm{geo}$ to see its impact for higher dimensional problems.
The set of Dirichlet boundary conditions can usually be described by simple geometries as well.
Hence, we denote $N_D$ the number of geometries describing the Dirichlet boundaries.
We approximate $\arccos\sqrt{\cdot}$ as in \cref{eq:arccos} truncated after $k=K$.

The algorithm starts by assigning three target registers for $H'_{ij}=H_{ij}/\norm{H}_\mathrm{max}$, $\sqrt{|H'_{ij}|}$, and $\vartheta_{ij}$ requiring $3r+1$ qubits next to the $2r$ input qubits.
Next, it tests if $x_i,x_j\in\mathcal{X}_\mathrm{D}$ and if $i\in\{1,N\}$.
This requires a total of $4 N_\mathrm{D}+2N_\mathrm{geo}$ comparisons combined, which requires a maximum of
\begin{equation}
	n_\mathrm{geo,a}
	=
	8 N_\mathrm{D}+4N_\mathrm{geo}-1
\end{equation}
ancilla qubits, including $4 N_\mathrm{D}+2N_\mathrm{geo}-2$ qubits for a multi-controlled Toffoli gate and one target qubit.
The runtime scales linearly with the number of comparisons, while each comparison contains two ADDs
\begin{equation}
	t_\mathrm{geo}
	=
	\bigO\left(
		\left(N_\mathrm{geo}+N_\mathrm{D}\right)t_{\mathrm{ADD}}
	\right).
\end{equation}
However, one can parallelize the comparisons for the additional cost of $\bigO((N_\mathrm{geo}+N_\mathrm{D})r)$ ancilla qubits.
Combining the results of the comparisons scales logaritmically with their number if executed in a cascade like structure.
With this we have
\begin{multline}
	t_{\mathrm{geo}}
	=
	\bigO\left(\log_2\left(N_\mathrm{D}+N_\mathrm{geo}\right)+t_{\mathrm{ADD}}\right)
	\\
	=
	\bigO\left(\log_2\left(N_\mathrm{geo}+N_\mathrm{D}\right)+r\right),
\end{multline}
assuming $t_\mathrm{ADD} = \bigO(r)$.

Based on those conditions, the renormalized contributions $H'_{ij}=H_{ij}/\norm{H}_\mathrm{max}$ are encoded using addition and the Kronecker-delta operation introduced in \cref{sec:arithmetic,sec:compare}, respectively.
This requires
\begin{equation}
	n_\mathrm{H',a}
	=
	r-1
\end{equation}
additional ancilla qubits for the Kronecker-delta including the target qubit.
The runtime is
\begin{equation}
	t_\mathrm{H'}
	=
	\bigO(\log_2 r+t_\mathrm{qcADD})
	=
	\bigO(r),
\end{equation}
where we assumed that $t_\mathrm{qcADD}=\bigO(r)$.

The transformation from two's complement into sign-magnitude form requires one controlled $\mathrm{qcSUB}_1$ followed by $r$ CNOT gates.
Neither of them requires ancilla qubits.
Their runtime is
\begin{equation}
	t_\mathrm{Trafo}
	=
	\bigO(r+t_\mathrm{qcADD})
	=
	\bigO(r).
\end{equation}
Here, we used that the runtime of a single CNOT gate is $t_\mathrm{CNOT}=\bigO(1)$.

The second last step is the computation of the square root in $\arccos\sqrt{\cdot}$ based on the Newton-Raphson method SQRT.
For this we need
\begin{equation}
	n'_\mathrm{SQRT,a}
	=
	(L+4)r +n_\mathrm{ADD,a}
\end{equation} ancilla qubits, the new target register included.
The runtime is
\begin{equation}
	t_\mathrm{SQRT}
	=
	\bigO(Lt_\mathrm{MUL}).
\end{equation}

The polynomial in \cref{eq:arccos} can be truncated after $k=K$ leading to.
\begin{equation}
	n'_\mathrm{POLY,a}
	=
	(K-1)(r+1)+n_\mathrm{ADD,a}
\end{equation}
additional ancilla qubits.
The runtime is
\begin{equation}
	t_\mathrm{POLY}
	=
	\bigO\left(
	Kt_\mathrm{MUL}
	\right)
	=
	\bigO\left(
	Kr^2
	\right).
\end{equation}
In the last step we assumed a carry-ripple based multiplier with $t_\mathrm{MUL}=\bigO(r^2)$.

In total, the oracle $O_\vartheta$ requires
\begin{equation}
	n_{O_\vartheta}
	= \bigO\left(
	(L+K+N_\mathrm{geo}+N_\mathrm{D})r
	\right)
\end{equation}
qubits and has a runtime of
\begin{equation}
	t_{O_\vartheta}
	=
	\bigO\left(
		(K+L)r^2+\log_2\left(N_\mathrm{geo}+N_\mathrm{D}\right)
	\right).
\end{equation}
Both, the memory requirements and the runtime are independent of $N$ and with it efficient.
One could argue that, $r$ has to be large enough to store values in $[0,N^2]$ for $i\in[N]$ and $1/\Delta^{2}=\bigO(N^2)$, which translates into $r=\bigO(\log_2N)$.
However, this makes the runtime only polylogarithmic in $N$ which is not endangering potential speed-ups in Ref.~\cite{danz2024calculating}.

\section{Conclusion}
\label{sec:conclusion}

In this work, we have provided a comprehensive discussion of the implementation of quantum oracles used in the block-encoding of matrices emerging in finite element analysis.
Starting from the basic quantum adder, and using fixed-point arithmetic, we have constructed explicit quantum circuits for the calculation of the necessary functions, which in the specific case require the evaluation of a polynomial combined with the square root.
In the appendix, we provided circuits for the implementation of logical operations such as comparison, conjunction and disjunction, that are also needed in this setting.
All these subroutines might be of general interest beyond the application to the construction of quantum oracles in \ac{fem}.

As a specific example, we focused on \ac{fem} for normal mode analysis of elastic structures, a problem that can be tackled in the quantum setting using quantum phase estimation, as discussed in Ref.~\cite{danz2024calculating}.
In this case, the problem is completely specified by the $N \times N$ mass $\bm{M}$ and stiffness $\bm{K}$ matrices, which we show how to obtain using finite element analysis in arbitrary dimensions and general geometries.
We also give them explicitly for the simple one-dimensional case.
Our scaling analysis reveals, that under reasonable assumptions on the approximation of the geometry, the construction of the oracles requires $\mathcal{O}(\mathrm{polylog}(N))$ ancilla qubits and similarly has $\mathcal{O}(\mathrm{polylog}(N))$ runtime.
This shows, that while the study of practical quantum advantages in \ac{fem} problems is still an open question, oracles should not be considered fundamental bottlenecks for potential polynomial or exponential quantum advantages from a complexity theoretic point of view.

\section*{Acknowledgements}
\label{sec:acknowledgements}
SD and TS were funded by the Bundesministerium für Wirtschaft und Klimaschutz (BMWK, Federal Ministry for Economic Affairs and Climate Action) in the quantum computing enhanced service ecosystem for simulation in manufacturing project (QUASIM, Grand No. 01MQ22001A).

\bibliography{lit.bib}
\appendix{}

\crefalias{section}{appendix}
\crefalias{subsection}{subappendix}

\section{Fixed-point exponentiation}
\label{sec:exponentiation}

In this appendix, we describe how to continue from the fixed-point multiplier achieving an exponentiatior allowing fractional exponents and with it how to compute signomial functions in a more general way than described in~\cref{sec:polynomials}.

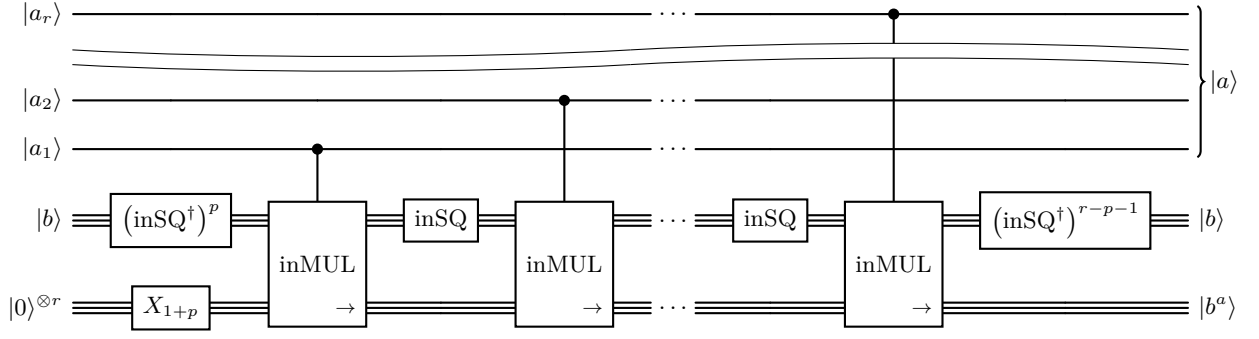
\begin{figure*}
	\centering
	\begin{adjustbox}{max width=\linewidth}
		\begin{quantikz}[wire types={q,q,q,q,b,b},classical gap=0.07cm]
			\lstick{$|a_r\rangle$}&&&&&\ \ldots\ &&\ctrl{4}&&\rstick[4]{$|a\rangle$}
			\\
			\wave &&&&&&&&&
			\\
			\lstick{$|a_2\rangle$} &&&&\ctrl{2}&\ \ldots\ &&&&
			\\
			\lstick{$|a_1\rangle$} &&\ctrl{1}&&&\ \ldots\ &&&&
			\\
			\lstick{$|b\rangle$} &\gate{\left(\mathrm{inSQ}^\dagger\right)^p}&\gate[2]{\mathrm{inMUL}}&\gate{\mathrm{inSQ}}&\gate[2]{\mathrm{inMUL}}& \ \ldots\ &\gate{\mathrm{inSQ}}&\gate[2]{\mathrm{inMUL}}&\gate{\left(\mathrm{inSQ}^\dagger\right)^{r-p-1}}&\rstick{$|b\rangle$}
			\\
			\lstick{$|0\rangle^{\otimes r}$} &\gate{X_{1+p}}&\gateoutput{$\rightarrow$}&&\gateoutput{$\rightarrow$}&\ \ldots\ &&\gateoutput{$\rightarrow$}&&\rstick{$|b^a\rangle$}
		\end{quantikz}
	\end{adjustbox}
	\caption[Circuit: quantum exponentiator]{
		Quantum exponentiatior (EXP) based on an arbitrary in-place quantum multiplier (inMUL) and an in-place square (inSQ).
		The $\mathrm{CNOT}^{\otimes r}$ denotes a series of $r$ parallel CNOT gates, with control on the qubits in $\ket{b}$ and applied to the qubits in the bottom register.
		The ancilla qubits for inMUL and inSQ are hidden.
	}
	\label{fig:exp}
\end{figure*}

\paragraph{Exponentiator}
Following the multiplier, the next operation is the quantum exponentiation EXP
\begin{equation}
	\mathrm{EXP}\ket{a,b}\ket{0}
	=
	\ket{a,b}\ket{b^a}.
\end{equation}
Here, we assume $a,b>0$ to prevent complex numbers if $a<1$.
Therefore, we require only $r$ qubits for $\ket{b}$ and $\ket{a}$ each (no sign qubit needed).
Similarly to \cref{eq:product}, the exponentiation can be decomposed into a product of multiples of $b$
\begin{equation}
	b^a
	=
	b^{\left(
			\sum_{k=0}^{r-1} 2^{k-p}a_{k+1}
			\right)}
	=
	\prod_{k=0}^{r-1} b^{\left(
			2^{k-p}a_{k+1}
			\right)}.
\end{equation}
The digit $a_{k+1}$ controls if the multiplier $b^{2^{k-p}}$ is part of the product or not.
In this case, we do not have a modified version of MUL that increases the input $b$ by the desired exponent $2^{k-p}$.
This means that we need another quantum routine that computes the square and square root of the input value $b$.
The full quantum exponentiator works as follows (see \cref{fig:exp}).
We start by reducing $b$ to $b^{2^{-p}}$ by applying the Hermitian conjugate of an in-place square operation inSQ $p$ times.
Parallel, we encode a $1$ in the target register by using a NOT gate $X$ applied to its $1+p$-th least significant qubit.
Next, the first contribution $b^{2^{-p}}$ for $k=0$ can be multiplied to the target register with an in-place multiplier controlled by $\ket{a_1}$.
After this, we square $b^{2^{-p}}$ once with inSQ achieving $|b^{2^{1-p}}\rangle$.
Another in-place multiplier, controlled by $\ket{a_2}$, multiplies the $b^{2^{1-p}}$ to the value in the target register.
The application of the square routine followed by a controlled multiplier will be repeated another $r-2$ times.
It is important that both the routine for the square and the multiplier are in-place routines.
Otherwise, we would require $\bigO(r^2)$ additional qubits to store intermediate products \cite{rieffel2014}.
In \cref{sec:in-place} we describe how to construct inSQ and inMUL.
This results in the state $|a,b^{2^{r-p-1}},b^a\rangle$.
The last step is to return the second register to $\ket{b}$.
This can be achieved by applying $\mathrm{inSQ}^\dagger$ $r-p-1$ times.

One can modify the given exponentiator for negative $b$ and integer $a$.
The routine requires only the addition of one Toffoli gate, applied to the sign qubit of the target register, at the end, while controlled by the sign qubit $\ket{b_r}$ and the smallest integer qubit $\ket{a_{p+1}}$.
The first $p$ inSQs would cancel with the $(\mathrm{inSQ}^\dagger)^p$ due to $\ket{a_p,\dots, a_1}=\ket{0}^{\otimes p}$.

Previous quantum-quantum exponentiators work similar to the routine described in this manuscript (cf. \cite{rieffel2014}).
However, we replaced out-of-place multiplier with in-place ones, reducing the memory requirements.
Furthermore, we describe exponentiation with fixed-point numbers allowing for non-integer exponents, which is useful for certain expansions (see \cref{sec:application}).

The memory requirements are dominated by the $n_\mathrm{inSQ}$ qubits of the in-place square operation (see \cref{sec:in-place,sec:newton-raphson}).
In addition, EXP needs $2r$ qubits.
Thus, for the total memory requirements $n_\mathrm{EXP}$ of the exponentiator and its ancilla requirements $n_\mathrm{EXP,a}$ we have
\begin{subequations}
	\begin{align}
		n_\mathrm{EXP}
		=
		(L+7)r+n_\mathrm{ADD,a}
		\\
		n_\mathrm{EXP,a}
		=
		(L+4)r+n_\mathrm{ADD,a}.
	\end{align}
\end{subequations}
Here, $L$ is the number of iterations required for inSQ (cf. \cref{sec:in-place,sec:newton-raphson}).
The runtime of the exponentiation consists of the contribution from the $2(r-1)$ repetitions of the inSQ and $\mathrm{inSQ}^\dagger$, and the $r$ in-place multiplications.
This gives a runtime $t_{\mathrm{EXP}}$ for the exponentiation of
\begin{equation}
	t_{\mathrm{EXP}}
	=
	2(r-1)t_{\mathrm{inSQ}}
	+rt_{\mathrm{inMUL}}
	=
	\bigO\left(
	L r t_{\mathrm{MUL}}
	\right),
\end{equation}
where $t_{\mathrm{inSQ}}$, and $t_{\mathrm{inMUL}}$ are the runtimes of the in-place square inSQ, and the in-place multiplier inMUL respectively.
For the second equation, we used that inSQ and inMUL contain both $\bigO(L)$ MULs (cf. \cref{sec:in-place,sec:newton-raphson}).

Quantum exponentiators are required for Shor's algorithm~\cite{shor1997}, where the base $b$ is known and only the exponent $a$ requires storage in a quantum register.
For this specific purpose, two quantum-classical exponentiators for integers, based on a carry-ripple adder, were already proposed in Refs.~\cite{vedral1996,beckman1996}.
They achieve an $\bigO(r^3)$ runtime with $7r+1$ and $5r+1$ qubits respectively (cf. \cref{tab:exp}).
\begin{table}
	\centering
	\caption{Quantum-classical exponentiation for $r$-digit exponents (required in Shor's algorithm)}
	\label{tab:exp}
	\begin{tabular}{lcc}
		\hline
		Method                                     & Runtime $t_\mathrm{EXP}$ & Memory $n_\mathrm{EXP}$
		\\
		                                           &                          & in qubits
		\\
		\hline
		Carry-ripple~\cite{beckman1996} & $\bigO(r^3)$             & $5r+3$
		\\
		Conditional sum~\cite{vanmeter2005}        & $\bigO(r\log_2^2 r)$     & $\approx 1.9r^2$
		\\
		Carry-lookahead~\cite{vanmeter2005}        & $\bigO(r\log_2^2 r)$     & $ \approx 1.9r^2$
		\\
		Fourier transform~\cite{beauregard2003}    & $\bigO(r^3)$             & $2r+3$
		\\
		Fourier transform~\cite{Pavlidis2013}      & $\bigO(r^2)$             & $9r+2$
		\\
		\hline
	\end{tabular}
\end{table}
Quantum-classical exponentiators, based on either the conditional sum or the carry-lookahead adder, both reach an $\bigO(r\log_2^2 r)$ runtime with $\approx 1.9r^2$ qubits~\cite{vanmeter2005}.
They achieve this speed-up by parallelizing the multiplier in blocks executed in a cascade-like structure.
A quantum-classical exponentiator based on \ac{qft} adders achieves an $\bigO(r^3)$ runtime in only $2r+3$ qubits.
This is described in Ref.~\cite{beauregard2003}.
At last, we also refer to Ref.~\cite{Pavlidis2013}, which contains a great overview of existing methods.
They, further propose a \ac{qft} based method that requires $9r+2$ qubits and has a runtime of $\bigO(r^2)$.
This is possible by merging classical-controlled rotation gates.
These results are summarized in \cref{tab:exp}.

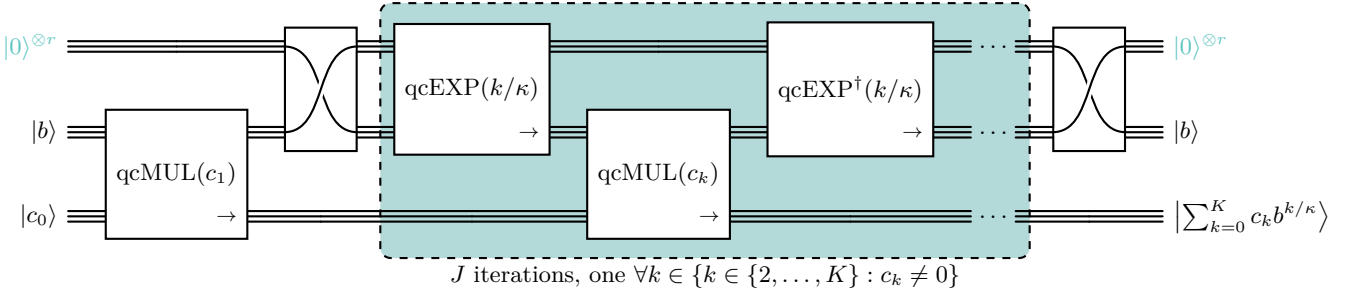
\begin{figure*}
	\centering
	\begin{adjustbox}{max width=\linewidth}
		\begin{quantikz}[wire types={b,b,b},classical gap=0.07cm]
			\lstick{$\blue{|0\rangle^{\otimes r}}$}&&\gate[2, swap]{}&\gate[2]{
				\mathrm{qcEXP}(k/\kappa)
			}\gategroup[
				3,
				steps=4,
				style={
						dashed,
						rounded corners,
						fill=teal!30,
						inner xsep=2pt
					},
				background,
				label style={
						label position=below,
						anchor=north,
						yshift=-0.15cm
					}
			]{$J$ iterations, one $\forall k\in\left\{k\in \{2,\dots,K\}:c_k\neq 0\right\}$}&&\gate[2]{
				\mathrm{qcEXP}^\dagger(k/\kappa)
			}&\ \ldots \ &\gate[2,swap]{}&\rstick{$\blue{|0\rangle^{\otimes r}}$}
			\\
			\lstick{$| b \rangle$}&\gate[2]{
				\mathrm{qcMUL}(c_1)
			}&&\gateoutput{$\rightarrow$}&\gate[2]{
				\mathrm{qcMUL}(c_k)
			}&\gateoutput{$\rightarrow$}& \ \ldots \ &&\rstick{$|b\rangle$}
			\\
			\lstick{$|c_0\rangle$}&\gateoutput{$\rightarrow$}&&&\gateoutput{$\rightarrow$}&& \ \ldots \ &&\rstick{$\left|\sum_{k=0}^K c_k b^{k/\kappa}\right\rangle$}
		\end{quantikz}
	\end{adjustbox}
	\caption{
		Quantum routine (SIG) for signomial functions based on quantum-classical versions of the multiplier (qcMUL) and exponentiation (qcEXP).
		The arguments in parentheses are the classical multipliers and exponents respectively.
		We hide the ancilla qubits required in qcMUL and qcEXP.
	}
	\label{fig:sig}
\end{figure*}

\paragraph{Operation for signomial functions}
At last, we propose a quantum routine for the computation of signomial functions of the form
\begin{equation}
	\mathrm{sig}(b)
	=
	\sum_{k=0}^{K}
	c_k b^{k/\kappa},
	\label{eq:sig}
\end{equation}
with $\kappa$ an integer and, by definition of a signomial, $b>0$.
Normally, signomial functions allow for arbitrary real exponents, but fixed-point arithmetic limits this to rational exponent $k/\kappa$.
The corresponding quantum routine SIG is defined by
\begin{equation}
	\mathrm{SIG}\ket{b}\ket{0}
	=
	\ket{b}\ket{\mathrm{sig}(b)}.
\end{equation}
Here, we assume that we know $K$, all $c_k$, and $\kappa$ in advance.
This allows us to use a quantum-classical exponentiator qcEXP in which $b$ remains in a quantum register but the exponent $a$ is stored classically\footnote{
	Not to confuse with the quantum-classical exponentiator in \cref{tab:exp}, where $a$ is stored in a quantum register and $b$ classically.
}, i.e. only operations controlled by a classical bit in the state $\ket{a_k}_c=\ket{1}_c$ remain.
This simplifies the quantum routine in width and depth.
Similarly, we can use a quantum-classical multiplier qcMUL for the multiplication with the classically stored weights $c_k$.
The full routine for the computation of the signomial starts with the initialization of $c_0$ in the target register after which we use a quantum-classical out-of-place multiplier $\mathrm{qcMUL}(c_1)$ to multiply $b$, stored in one quantum register, with $c_1$, stored in a classical register, and add the product to the target register (see \cref{fig:sig}).
After this a series of $J$ similar iterations follows, where $J$ is the number of remaining summands with non-zero weight $c_k,\forall k\geq 2$.
They all start with increasing the power of $b$ with the quantum-classical exponentiator $\mathrm{qcEXP}(k/\kappa)$ to $b^{k/\kappa}$, followed by a quantum-classical multiplier with the corresponding weight $c_k$ adding their product $c_k b^{k/\kappa}$ to the target register.
The last part of one iteration is to return $\ket{b^{k/\kappa}}\to \ket{b}$ with the inverse $\mathrm{qcEXP}^\dagger(k/\kappa)$.
The reinitialization of $\ket{b}$ at the end of each step can be combined with the following exponentiation going directly to the next power (e.g. $|b^2\rangle\to|b^5\rangle$) by slightly modifying qcEXP.
Even though the combined reinitialization reduces the runtime of the routine, it is not changing the overall scaling which is why we consider the previously described routine in the following.

The routine for a signomial with $J$ non-zero weights consists of $\bigO( J)$ quantum-classical multiplier and $\bigO(J)$ quantum-classical exponentiator in series yielding a runtime of
\begin{equation}
	t_{\mathrm{SIG}}
	=
	\bigO\left(
	J(
	t_{\mathrm{qcMUL}}
	+t_{\mathrm{qcEXP}}
	)
	\right).
	\label{eq:lat_sig1}
\end{equation}
Here, $t_{\mathrm{qcMUL}}$ and $t_{\mathrm{qcEXP}}$ are the runtimes of the used multiplier and exponentiator, respectively.
The advantage of quantum-classical multiplication and exponentiation is that the number of internal steps is at most $\bar{r}$ instead of $r$, where $\bar{r}$ is the number of bits required to represent the maximum factor or exponent respectively.
This reduces the runtime of qcMUL and qcEXP to
\begin{subequations}
	\begin{align}
		t_\mathrm{qcMUL}
		 & =
		\bigO(
		t_\mathrm{ADD}\log_2 \left(
			\abs{c}_\mathrm{max}
			\right)
		),
		\\
		t_\mathrm{qcEXP}
		 & =
		\bigO(
		t_\mathrm{MUL}L\log_2 (K\kappa)
		),
	\end{align}
\end{subequations}
where $\log_2\abs{c}_\mathrm{max}=\max_k \log_2[\lfloor\abs{c_k}\rfloor(\abs{c_k}-\lfloor\abs{c_k}\rfloor)]$ is the maximum number of digits required for the weights $c_k$ occurring in \cref{eq:sig}.
Analogously, $\log_2 (K\kappa)$ digits are required to represent the exponent $K/\kappa$.
This reduces the runtime (\cref{eq:lat_sig1}) to
\begin{equation}
	t_\mathrm{SIG}
	=
	\bigO(
	JL \log_2 (K\kappa)t_\mathrm{MUL}
	).
	\label{eq:runtime_sig}
\end{equation}
Here, we assume $\log_2\abs{c}_\mathrm{max}< r$ and with it $t_\mathrm{MUL}>\log_2{\abs{c}_\mathrm{max}}t_\mathrm{ADD}$, i.e. $t_\mathrm{qcMUL}< t_\mathrm{qcEXP}$.
The computation of a signomial function requires a total of $n_\mathrm{SIG}$ qubits containing $n_\mathrm{SIG,a}$ ancillas
\begin{subequations}
	\begin{align}
		n_\mathrm{SIG}
		 & =
		r+n_\mathrm{qcEXP}
		=
		(L+7)r +n_\mathrm{ADD,a},
		\\
		n_\mathrm{SIG,a}
		 & =
		(L+5)r +n_\mathrm{ADD,a}.
	\end{align}
\end{subequations}
Here, $n_\mathrm{qcEXP}=n_\mathrm{EXP}-r$ is the number of qubits required for the quantum-classical exponentiator that is not controlled by a second quantum register.

One can modify this routine for multi-variable signomials $\mathrm{sig}(a,b,\dots)$ by combining quantum-classical exponentiators qcEXP that sample from different quantum registers with the various bases ($\ket{a}$, $\ket{b}$, \dots).
The different powers of $a$, $b$, \dots can be combined with quantum-quantum in-place multipliers inMUL before they will be added to the target register with a qcMUL.
This is not affecting the runtime, but increases the memory requirements by $Mr$ qubits, where $M$ is the number of variables in the signomial.
They consist of $M-1$ additional input registers and $1$ ancilla register to store intermediate products.

\section{In-place arithmetic}
\label{sec:in-place}
Some quantum implementations of arithmetic operations like the multiplier MUL or exponentiator EXP are naturally out-of-place.
Based on numbers, encoded in the base of two input register, they prepare another number in a third register.
This yields several wasted quantum registers if one stacks multiple of those operations in series.
We can prevent this by using in-place versions instead, that reuse one of the input registers to store the result.
The structure of in-place operations (based on their out-of-place versions) is always the same and can be split into three parts~\cite{vedral1996} as shown in the following for the computation of an invertible function $f(a)$:
\begin{enumerate}
	\item Computation of $f(a)$ with the out-of-place arithmetic operator $U_f$
	      \begin{equation}
		      U_f \ket{a,0}
		      =
		      \ket{a,f(a)}.
	      \end{equation}
	\item Swap of the target and input register
	      \begin{equation}
		      \mathrm{SWAP}\ket{a,f(a)}
		      =
		      \ket{f(a),a}.
	      \end{equation}
	\item Annihilation of the input $a$ with the Hermitian conjugate $U^\dagger_{f^{-1}}$:
	      \begin{equation}
		      U^\dagger_{f^{-1}} \ket{f(a),a}
		      =
		      \ket{f(a),0}.
	      \end{equation}
\end{enumerate}
In this section, we describe the two in-place operations inSQ and inMUL, occurring in the exponentiator in \cref{sec:polynomials}.

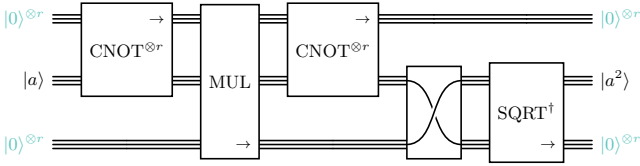
\begin{figure}
	\centering
	\begin{adjustbox}{max width=\linewidth}
		\begin{quantikz}[wire types={b,b,b},classical gap=0.07cm]
			\lstick{$\blue{|0\rangle^{\otimes r}}$} &\gate[2]{\mathrm{CNOT}^{\otimes r}}\gateoutput{$\rightarrow$}&\gate[3]{\mathrm{MUL}}&\gate[2]{\mathrm{CNOT}^{\otimes r}}\gateoutput{$\rightarrow$}&&&\rstick{$\blue{|0\rangle^{\otimes r}}$}
			\\
			\lstick{$|a\rangle$} &&&&\gate[2,swap]{}& \gate[2]{\mathrm{SQRT}^\dagger}&\rstick{$|a^2\rangle$}
			\\
			\lstick{$\blue{|0\rangle^{\otimes r}}$} &&\gateoutput{$\rightarrow$}&&&\gateoutput{$\rightarrow$}&\rstick{$\blue{|0\rangle^{\otimes r}}$}
		\end{quantikz}
	\end{adjustbox}
	\caption{
		Quantum in-place square (inSQ) based on a quantum multiplier (MUL) and a square root (SQRT) routine.
		The $\mathrm{CNOT}^{\otimes r}$ represents $r$ parallel CNOT gates.
		The ancilla qubits used in MUL and SQRT are hidden.
	}
	\label{fig:square}
\end{figure}

\paragraph{Square (inSQ)}
The quantum implementation for the in-place square operation inSQ is shown in \cref{fig:square}.
It starts with the computation of the square.
For this it copies the input value $a$ to an ancilla register by utilizing a column of CNOT gates, before it uses both registers as input for the out-of-place multiplication MUL.
The product $a^2$ is stored in another ancilla register.
The first ancilla register, storing $\ket{a}$ can now be reinitialized.
For the second step, we swap the two register $\ket{a}$ and $|a^2\rangle$.
At last, we need the inverse of the original arithmetic operation which is the square root.
By using the register that stores $|a^2\rangle$, we can compute $a$ without accessing the $\ket{a}$ register.
The Hermitian conjugate of the corresponding quantum gate allows us to reset $a$ to $\ket{0}^{\otimes r}$ in the ancilla register.

The memory requirements consist of one $r$-qubit ancilla register next to the $n_\mathrm{SQRT}$ qubits necessary for SQRT (see \cref{sec:newton-raphson}).
This adds up to
\begin{subequations}
	\begin{align}
		n_\mathrm{inSQ}
		 &
		=
		(L+3)r+n_\mathrm{ADD,a},
		\\
		n_\mathrm{inSQ,a}
		 &
		=
		(L+2)r+n_\mathrm{ADD,a}.
	\end{align}
\end{subequations}
$n_\mathrm{inSQ,a}$ denotes the ancilla qubits required for inSQ.
The runtime $t_\mathrm{inSQ}$ is dominated by the runtime $t_\mathrm{SQRT}$ of SQRT
\begin{equation}
	t_\mathrm{inSQ}
	=
	\bigO(
	t_\mathrm{SQRT}
	)
	=
	\bigO(
	Lt_\mathrm{MUL}
	).
\end{equation}

\begin{figure}
	\centering
	\begin{adjustbox}{max width=\linewidth}
		\begin{quantikz}[wire types={b,b,b,b},classical gap=0.07cm]
			\lstick{$\blue{|0\rangle^{\otimes r}}$} &&\gate[2]{\mathrm{REC}}\gateoutput{$\rightarrow$}&\gate[2,swap]{}&&\gate[2,swap]{}&\gate[2]{\mathrm{REC}^\dagger}\gateoutput{$\rightarrow$}&\rstick{$\blue{|0\rangle^{\otimes r}}$}
			\\
			\lstick{$|a\rangle$} &\gate[3]{\mathrm{MUL}}&&&\gate[3]{\mathrm{MUL}^\dagger}&&&\rstick{$|a\rangle$}
			\\
			\lstick{$|b\rangle$} &&\gate[2,swap]{}&&&&&\rstick{$|ab\rangle$}
			\\
			\lstick{$\blue{|0\rangle^{\otimes r}}$} &\gateoutput{$\rightarrow$}&&&\gateoutput{$\rightarrow$}&&&\rstick{$\blue{|0\rangle^{\otimes r}}$}
		\end{quantikz}
	\end{adjustbox}
	\caption{
		Quantum in-place multiplier (inMUL) based on an arbitrary quantum multiplier (MUL) and an operation (REC) that computes the reciprocal.
		The ancilla qubits necessary for MUL and REC are not shown.
	}
	\label{fig:in_mult}
\end{figure}
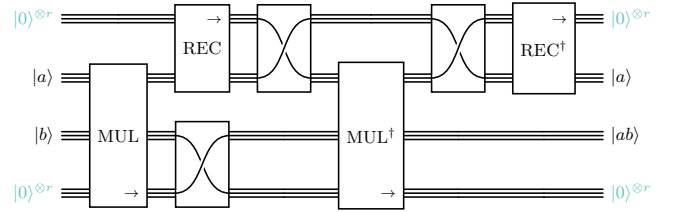

\paragraph{Multiplication (inMUL)}
The in-place multiplication inMUL starts similar with the computation of the product $ab$ via MUL storing it in an ancilla register (cf. \cref{fig:in_mult}).
The second step is the swap of one of the inputs ($\ket{b}$) with the target register storing $\ket{ab}$.
The last step requires a quantum divider, which is equal to the multiplication of the reciprocal of the divisor with the dividend
\begin{equation}
	\frac{\mathrm{dividend}}{\mathrm{divisor}}
	=
	\mathrm{dividend} \times \mathrm{divisor}^{-1}.
\end{equation}
Hence, we start by computing the reciprocal of $a$ with REC and store it in an ancilla register.
Next, we multiply $a^{-1}$ with the previous computed product $ab$ resulting in $b$.
The Hermitian conjugate of MUL annihilates $b$ in the ancilla register.
It remains to undo REC.

This routine requires a total memory of
\begin{subequations}
	\begin{align}
		n_\mathrm{inMUL}
		 & =
		2r+n_\mathrm{REC}
		=
		(L+5)r+n_\mathrm{ADD,a}
		\\
		n_\mathrm{inMUL,a}
		 & =
		(L+3)r+n_\mathrm{ADD,a},
	\end{align}
\end{subequations}
with $n_\mathrm{inMUL}$ the total qubit number, $n_\mathrm{inMUL,a}$ the ancillas, and $n_\mathrm{REC}$ is the memory requirement of REC (cf. \cref{sec:reciprocal}).
The runtime $t_\mathrm{inMUL}$ is dominated by the runtime $t_\mathrm{REC}$ of REC:
\begin{equation}
	t_\mathrm{inMUL}
	=
	\bigO(
	t_\mathrm{MUL}
	+t_\mathrm{REC}
	)
	=
	\bigO(
	Lt_\mathrm{MUL}
	).
\end{equation}

The in-place variants described here require additional operations (REC and SQRT).
Those can be implemented using the Newton-Raphson method, that is described in \cref{sec:newton-raphson,sec:reciprocal}.

\section{Newton-Raphson Reciprocal}
\label{sec:reciprocal}

Similarly to the square root in \cref{sec:newton-raphson} we can use the Newton-Raphson method to compute the reciprocal.
For this, we require a function with a root at $x=R^{-1}$, where $R$ is the input value.
A common choice is
\begin{equation}
	f_R(x)
	=
	\frac{1}{x}-R,
\end{equation}
which, from \cref{eq:nr_iteration}, gives
\begin{equation}
	x_{k+1}
	=
	x_k\left(2-Rx_k\right).
\end{equation}
This converges with quadratic order towards $R^{-1}$, which we can show by defining an error $\varepsilon_k=\abs{1-Rx_k}$ and study how it changes with increasing $k$
\begin{equation}
	\varepsilon_{k+1}
	=
	\abs{1-Rx_{k+1}}
	=
	\left(1-Rx_{k}\right)^2
	=
	\varepsilon_k^2.
\end{equation}
With this, the number of correct digits in $x_k$ doubles with every iteration, if $Rx_0\in(0,2)$ (cf. Refs.~\cite{flynn1970,parker1992}).
For the initial estimation $x_0$ of the reciprocal we can rely on a method, that is based on finding the leading ``1'' in the binary form and moving it to the other side of the binary point as described in Ref.~\cite{Haner2018}.
Thies gives an approximation that is good enough to require only a few Newton-Raphson iterations (2-4 in their case).

\begin{figure*}
	\centering
	\begin{adjustbox}{max width=\linewidth}
		\begin{quantikz}[transparent,wire types={b,b,b,b,b},classical gap=0.07cm]
			\lstick{$\ket{R}$} &\gate[3]{\mathrm{MUL}} &&&&&&\gate[3]{\mathrm{MUL}^\dagger}&\rstick{$\ket{R}$}
			\\
			\lstick{$\ket{x_k}$} &&&\gate[2,swap]{}&&\gate[2,swap]{}&&&\rstick{$\ket{x_k}$}
			\\
			\lstick{$\blue{\ket{0}^{\otimes r}}$} &\gateoutput{$\rightarrow$}&\gate[2]{\mathrm{SUB}}&&\gate[3]{\mathrm{MUL}} &&\gate[2]{\mathrm{ADD}}&\gateoutput{$\rightarrow$}&\rstick{$\blue{\ket{0}^{\otimes r}}$}
			\\
			\lstick{$\blue{\ket{0}^{\otimes r}}$}&\gate{X_{2+p}}&\gateoutput{$\rightarrow$}&&&&\gateoutput{$\rightarrow$}&\gate{X_{2+p}}&\rstick{$\blue{\ket{0}^{\otimes r}}$}
			\\
			\lstick{$\ket{0}^{\otimes r}$} &&&&\gateoutput{$\rightarrow$}&&&&\rstick{$\ket{x_{k+1}=x_{k}(2-Rx_k)}$}
		\end{quantikz}
	\end{adjustbox}
	\caption{
		Circuit for one iteration of the quantum Newton-Raphson reciprocal $R^{-1}$.
		The ancilla qubits necessary for MUL, SUB and ADD are not shown.
	}
	\label{fig:nr_rec}
\end{figure*}
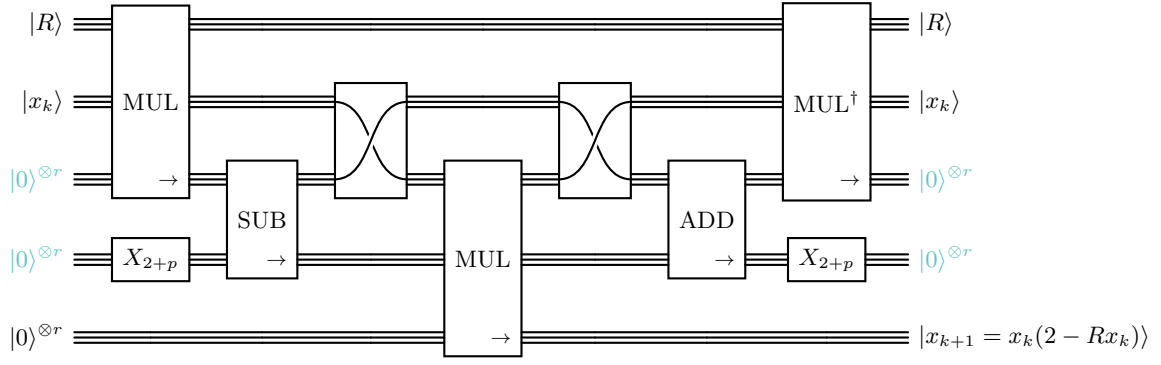

The corresponding quantum circuit consists of multiplication and subtraction (cf. \cref{fig:nr_rec}).
Each iterative step starts with the multiplication of $x_k$ and $R$.
The result gets subtracted from another register in which we encoded the number 2 with a single not gate applied to the $(2+p)$-th least significant qubit, where $p$ is the number of digits after the binary point.
A multiplication of the difference $(2-Rx_k)$ with the initial estimation $x_k$ follows, giving our next estimation $x_k$ stored in the lowest register of \cref{fig:nr_rec}.
It remains to uncompute the previous steps reinitializing the ancilla registers.
This gives on step of the Newton-Raphson method.
The full series is analogous to \cref{fig:nr}.

One iteration of the Newton-Raphson step for the reciprocal requires
\begin{subequations}
	\begin{align}
		n_\mathrm{REC-step}
		 & =
		5r+n_\mathrm{ADD,a},
		\\
		n_\mathrm{REC-step,a}
		& =
		3r+n_\mathrm{ADD,a},
	\end{align}
\end{subequations}
qubits and the full circuit with $L$ steps
\begin{subequations}
	\begin{align}
		n_\mathrm{REC}
		 & =
		(L+3)r+n_\mathrm{ADD,a},
		\\
		n_\mathrm{REC,a}
		& =
		(L+1)r+n_\mathrm{ADD,a},
	\end{align}
\end{subequations}
qubits.
The runtime of the full circuit is
\begin{multline}
	t_\mathrm{REC}
	=
	L\left(
	3t_\mathrm{MUL}
	+2t_\mathrm{ADD}
	+2t_\mathrm{SWAP}
	\right)
	+ 2t_{x_0}
	\\
	=
	\bigO(
	Lt_\mathrm{MUL}
	).
\end{multline}
Here, $t_\mathrm{SWAP}$ is the runtime of one (or $r$ parallel) SWAP gates, which is negligible compared to $t_\mathrm{MUL}$.
The runtime $t_{x_0}$ of the encoding of $x_0$ is also small compared with $t_\mathrm{MUL}$.

\section{Quantum logic operators}
\label{sec:logic_operator}

The piece-wise functions necessary for computing the elements of $\bm{H}$ (see \cref{eq:H_1d}) require the computation of boolean conditions.
In this appendix, we propose quantum implementations for two types of operations:
\begin{itemize}
	\item Comparisons like \emph{equal} ($=$) and \emph{greater than} ($>$).
	\item Boolean logical operations like \emph{conjunction} ($\land$) and \emph{disjunction} ($\lor$).
\end{itemize}

\subsection{Comparisons}
\label{sec:compare}
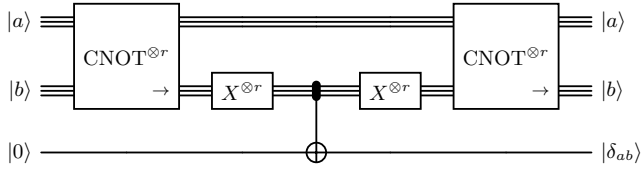
\begin{figure}
	\centering
	\begin{adjustbox}{max width=\linewidth}
		\begin{quantikz}[wire types={b,b,q},classical gap=0.07cm]
			\lstick{$|a\rangle$} &\gate[2]{\mathrm{CNOT}^{\otimes r}}&&&&\gate[2]{\mathrm{CNOT}^{\otimes r}} &\rstick{$|a\rangle$}
			\\
			\lstick{$|b\rangle$} &\gateoutput{$\rightarrow$}&\gate{X^{\otimes r}}&\ctrl{1}&\gate{X^{\otimes r}}&\gateoutput{$\rightarrow$}&\rstick{$|b\rangle$}
			\\
			\lstick{$|0\rangle$} &&&\targ{}&&&\rstick{$|\delta_{ab}\rangle$}
		\end{quantikz}
	\end{adjustbox}
	\caption[Circuit: Kronecker delta]{
		Quantum circuit for the Kronecker delta operation for two $r$-qubit registers $\ket{a}$ and $\ket{b}$.
		The $\mathrm{CNOT}^{\otimes r}$ represents $r$ parallel CNOT gates while the center gate is an $r$-qubit controlled NOT gate with $r-2$ hidden ancilla qubits.
	}
	\label{fig:kronecker}
\end{figure}

In order to compare two binary numbers ($a$ and $b$) stored in the computational basis of two quantum registers, we can make use of CNOT gates.
For two registers of $r$ qubits each (focusing on positive numbers, $a,b>0$), we apply a series of $r$ CNOT gates between qubit pairs of equal significance (see \cref{fig:kronecker}).
If both numbers are equal this will fully annihilate any $1$ in the target register.
One application of NOT gates to all qubits of this register inverts all of them leaving $1$s in all qubits if $a=b$.
An $r$-qubit controlled Toffoli gate targeting an ancilla qubit follows.
This triggers only if all control qubits are 1 and with it if $a=b$.
Hence, we have the Kronecker-delta $\delta_{ab}$ in the target register.
It remains to restore the initial values $a$ and $b$.

This routine requires a total of
\begin{equation}
	n_\mathrm{\delta}
	=
	2r+1+n_{r\mathrm{TOF},a}
	=3r-1,
\end{equation}
qubits.
Here, $n_{r\mathrm{TOF},a}=r-2$ is the ancilla number of the $r$-qubit controlled Toffoli gate~\cite{barenco1995}.
The runtime is equal to
\begin{equation}
	t_\mathrm{\delta}
	=4+t_{r\mathrm{TOF}}
	=\bigO\left(t_\mathrm{TOF} \log_2 r \right)
	=\bigO(\log_2 r),
\end{equation}
where we introduced the runtime of the $r$-qubit Toffoli gate $t_{r\mathrm{TOF}}$ measured in Toffoli gates $t_\mathrm{TOF}=\bigO(1)$ executed parallel in a cascade like structure.

\begin{figure}
	\hspace{-1cm}
	\begin{adjustbox}{max width=\textwidth}
		\begin{quantikz}[wire types={b,b,b,q},classical gap=0.07cm]
			\lstick{$|a\rangle$} &\gate[4,label style={yshift=5mm}]{\mathrm{SUB}} &\gate[3]{\mathrm{ADD}} &\rstick{$|a\rangle$}
			\\
			\lstick{$\blue{|0\rangle}$} &&&\rstick{$\blue{|0\rangle}$}
			\\
			\lstick{$|b\rangle$} &\gateoutput{$\rightarrow$} &\gateoutput{$\rightarrow$} &\rstick{$|b\rangle$}
			\\
			\lstick{$|0\rangle$} &\gateoutput{$\rightarrow$} &\rstick{$|\sgnb{b-a}\rangle=|a>b\rangle$}
		\end{quantikz}
	\end{adjustbox}
	\begin{tikzpicture}[thick,remember picture,overlay]
		\node[inner sep=0pt] (circle) at (-0.1,0.6)
		{\includegraphics[width=2.7cm]{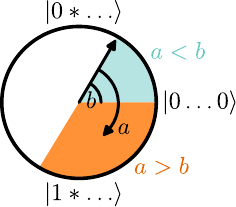}};
	\end{tikzpicture}
	\caption{Quantum circuit for the \emph{greater than} operation ($>$).}
	\label{fig:gr_le}
\end{figure}

The previous methods allow us to test if $a=b$, but what if we are interested in $a>b$ or $a<b$?.
For this we propose another routine based on ADD and SUB.
We start by increasing the $r$-qubit register storing $b$ by one qubit as the new most-significant bit, i.e., $\ket{b}\to \ket{0,b}$.
Then we subtract $a$ from $b$ using an $r+1$ qubit SUB applied to $\ket{a}\ket{0,b}$.
\begin{equation}
	\mathrm{SUB}\ket{a}\ket{0,b}
	=
	\ket{a}\ket{b-a\mod 2^{r+1}}.
\end{equation}
This is shown in \cref{fig:gr_le}.
Here, it is important that subtraction in the two's complement form  can lead to negative numbers indicated by a $1$ in the most significant qubit, i.e. only if $a>b$, the most significant qubit changes from $\ket{0}\to\ket{1}$
\begin{multline}
	\ket{b-a\mod 2^{r+1}}
	\\
	=
	\ket{\sgnb{b-a}}\ket{b-a\mod 2^r}.
\end{multline}
At last, we need to restore $b$ by applying an $r$ qubit ADD to the two input registers only.

This routine requires
\begin{equation}
	n_{>}
	=
	2r+1+n_\mathrm{ADD,a}
\end{equation}
qubits, where $n_\mathrm{ADD,a}$ is the number of qubits necessary for an $r+1$ qubit adder.
We need one qubit less because $\ket{a}$ is only an $r$-qubit register.
The runtime is governed by the adder
\begin{equation}
	t_{>}
	=
	\bigO(t_\mathrm{ADD}).
\end{equation}

Both comparisons ($(=)$ and $(>)$) can be extended for negative numbers by adding one qubit to the registers and using the two's complement form.

\subsection{Boolean logical operations}
\label{sec:truth}
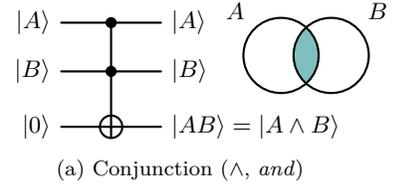
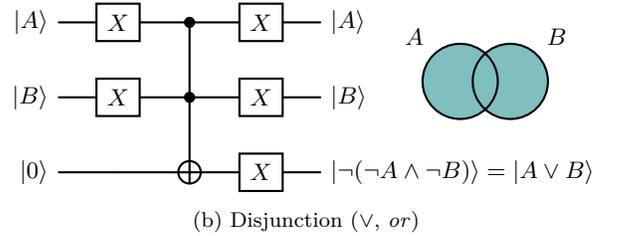
\begin{figure}
	\subfloat[Conjunction ($\land$, \emph{and})]{\label{fig:conjunction}
		\begin{adjustbox}{max width=\linewidth}
			\begin{quantikz}[wire types={q,q,q}]
				\lstick{$|A\rangle$}  &\ctrl{1} &\rstick{$|A\rangle$}
				\\
				\lstick{$|B\rangle$} &\ctrl{1} &\rstick{$|B\rangle$}
				\\
				\lstick{$|0\rangle$} &\targ{} &\rstick{$|A B\rangle=|A\land B\rangle$}
			\end{quantikz}
		\end{adjustbox}
		\begin{tikzpicture}[thick,remember picture,overlay,
				set/.style = { circle, minimum size = 1cm}]

			\node[set,label={135:$A$}] (A) at (-1,0.3) {};

			\node[set,label={45:$B$}] (B) at (-1/3,0.3) {};
			\begin{scope}
				\clip (-1,0.3) circle(0.5cm);
				\clip (-1/3,0.3) circle(0.5cm);
				\fill[teal!50](-1,0.3) circle(0.5cm);
			\end{scope}

			\draw (-1,0.3) circle(0.5cm);
			\draw (-1/3,0.3) circle(0.5cm);
		\end{tikzpicture}
	}

	\subfloat[Disjunction ($\lor$, \emph{or})]{\label{fig:disjunction}
		\centering
		\begin{adjustbox}{max width=\linewidth}
			\begin{quantikz}[wire types={q,q,q}]
				\lstick{$|A\rangle$}  &\gate{X}&\ctrl{1}&\gate{X}&\rstick{$|A\rangle$}
				\\
				\lstick{$|B\rangle$} &\gate{X}&\ctrl{1} &\gate{X}&\rstick{$|B\rangle$}
				\\
				\lstick{$|0\rangle$} &&\targ{} &\gate{X}&\rstick{$|\lnot(\lnot A\land \lnot B)\rangle=|A\lor B\rangle$}
			\end{quantikz}
		\end{adjustbox}
		\begin{tikzpicture}[thick,remember picture,overlay,
				set/.style = { circle, minimum size = 1cm}]

			\node[set,fill=teal!50,label={135:$A$}] (A) at (-2.,0.3) {};

			\node[set,fill=teal!50,label={45:$B$}] (B) at (-4/3,0.3) {};

			\draw (-2,0.3) circle(0.5cm);
			\draw (-4/3,0.3) circle(0.5cm);

		\end{tikzpicture}
	}
	\caption{
		Quantum circuit and Venn diagrams for the conjunction ($\land$, \emph{and}) and the  disjunction ($\lor$, \emph{or}) operation in (a) and (b), respectively.
		$A$ and $B$ are boolean variables being either $0$ or $1$ for \texttt{false} and \texttt{true}.}
	\label{fig:junction}
\end{figure}

\cref{eq:H_1d} requires the combination of multiple comparisons.
This can be achieved with the conjunction ($\land$, \emph{and}).
The Toffoli gate is the natural quantum implementation of the conjunction as it inverts the information in the target qubit only if both control qubits $\ket{A}$ and $\ket{B}$ are $\ket{1}$ (\texttt{true}).
Applied to an ancilla qubit in state $\ket{0}$, this yields in $\ket{0}$ (\texttt{false}) for $AB=0$ and $\ket{1}$ (\texttt{true}) for $AB=1$ (see \cref{fig:conjunction}).
Multiple Toffoli gates can be combined for more than two conditions (cf. $r$-qubit Toffoli gate~\cite{barenco1995}).

Complex geometries require the combination of conditions with the disjunction ($\lor$, \emph{or}).
For this we use the logical not ($\neg$) and invert the two conditions $A$ and $B$ before we combine them with a Toffoli.
This gives $\neg A\land \neg B$, which is the opposite of what we desire.
Another negation results in
\begin{equation}
	\neg(\neg A \land \neg B)
	=
	A\lor B.
\end{equation}
Translated in quantum operations we need to invert the input and output of a Toffoli gate with NOT gates (see \cref{fig:disjunction}).

\section{Advanced contours}
\label{sec:adv_contours}

In \cref{sec:contour} we already described how to test if a node is within a geometry consisting purely of hypercuboids.
In this appendix, we extend the list of primitive geometries and give an idea of how to construct more advanced geometries.
The second primitive is a $d$-dimensional hyperellipsoid $\mathcal{E}^d$ limited by
\begin{equation}
	\sum_{\alpha=1}^d
	\left(
	\frac{\Delta i^{(\alpha)}-c^{(\alpha)}}{h^{(\alpha)}}
	\right)^2
	\leq 1,
\end{equation}
where $\bm{c}=(c^{(1)},\dots,c^{(d)})^T$ is the center of the hyperellipsoid and $h^{(\alpha)}$ is the length of its semi-axes in direction $\alpha$.
The implementation of this condition requires the computation of a polynomial with $d$ terms (see \cref{sec:polynomials}) and the comparison with $1$ (see \cref{sec:compare}).

The conditions for hypercuboids and hyperellipsoids can be combined partially into other geometries like a cylinder
\begin{multline}
	\Delta^{-1}a^{(1)}
	\leq
	i^{(1)}
	\leq
	\Delta^{-1}b^{(1)}
	\\
	\land
	\sum_{\alpha=2}^3
	\left(
	\frac{\Delta i^{(\alpha)}-c^{(\alpha)}}{h^{(\alpha)}}
	\right)^2
	\leq
	1,
\end{multline}
or the geometries can be combined into bigger geometries (e.g. $\bm{x}_i \in \mathcal{C}^d \cup \mathcal{E}^d$).
Rotated versions of those geometries can be used by rotating the position variable (e.g. $x^{(1)}\to \cos \theta x^{(1)}+\sin \theta x^{(2)}$).
The list of conditions can be extended by arbitrary signomial functions describing the limits of the geometry if necessary.

\end{document}